\documentstyle[psfig,macros, amsmath,fleqn]{mn}
\begin{document}
%

\title[Statistics of Dark Matter Substructure III]
      {Statistics of Dark Matter Substructure:  
       III. Halo-to-Halo Variance}

\author[Jiang \& van den Bosch]
       {Fangzhou Jiang\thanks{E-mail:fangzhou.jiang@yale.edu} 
        \& Frank C. van den Bosch \\
        Department of Astronomy, Yale University, New Haven, CT 06511, USA}


\date{}

\pagerange{\pageref{firstpage}--\pageref{lastpage}}
\pubyear{2016}

\maketitle

\label{firstpage}


\begin{abstract}
  We present a study of unprecedented statistical power regarding the
  halo-to-halo variance of dark matter substructure. Using a
  combination of $N$-body simulations and a semi-analytical model, we
  investigate the variance in subhalo mass fractions and subhalo
  occupation numbers, with an emphasis on how these statistics scale
  with halo formation time.  We demonstrate that the subhalo mass
  fraction, $\fsub$, is mainly a function of halo formation time, with
  earlier forming haloes having less substructure. At fixed formation
  redshift, the average subhalo mass fraction is virtually independent
  of halo mass, and the mass dependence of $\fsub$ is therefore mainly
  a manifestation of more massive haloes assembling later.  We compare
  observational constraints on $\fsub$ from gravitational lensing to
  our model predictions and simulation results. Although the inferred
  $\fsub$ are substantially higher than the median $\Lambda$CDM
  predictions, they fall within the 95th percentile due to
  halo-to-halo variance. We show that the halo occupation
  distributions of subhaloes do not follow Poisson statistics; whereas
  $P(N|M)$ is super-Poissonian for large $\Nave$, a result that is
  well established, it becomes sub-Poissonian for $\Nave \la 2$. We
  show that ignoring this non-Poissonity results in systematic errors
  of the predicted clustering of galaxies of a few percent, and with a
  complicated scale- and luminosity-dependence. Haloes that assemble
  earlier have $P(N|M)$ that are closer to a Poisson distribution,
  suggesting that the dynamical evolution of subhaloes drives the
  statistics towards Poissonian. However, contrary to a recent claim,
  the non-Poissonity of subhalo occupation statistics does not vanish
  by selecting haloes with fixed mass and fixed formation redshift.
  Finally, we use subhalo occupation statistics to put loose
  constraints on the mass and formation redshift, $\zf$, of the Milky
  Way halo. Using observational constraints on the maximum circular
  velocities of the three most massive satellite galaxies in the Milky
  Way, we infer that $0.25<\Mvir/10^{12}\Msunh<1.4$ and $0.1<\zf<1.4$
  at 90\% confidence level.
\end{abstract} 


\begin{keywords}
methods: analytical --- 
methods: statistical --- 
galaxies: haloes --- 
dark matter
\end{keywords}



\section{Introduction} 
\label{Sec:Introduction}

In the hierarchical $\Lambda$+CDM structure formation paradigm, dark
matter haloes contain subhaloes, which are the remnants of halos that
have been accreted by their host halo over cosmic time, and have
survived tidal destruction.  Since the assembly history of a dark
matter halo depends on halo mass, cosmology, and its large scale
environment, so do the statistics of subhaloes. As a consequence, the
large halo-to-halo variance in halo assembly histories gives rise to a
very significant halo-to-halo variance in the substructure content of
dark matter haloes, even at a fixed halo mass.

An accurate knowledge of this halo-to-halo variance has numerous
applications.  For example, the severity of the `too-big-to-fail'
problem (Boylan-Kolchin \etal 2011) strongly depends on the
halo-to-halo variance of the abundance and internal densities of
massive subhaloes (e.g., Purcell \& Zentner 2012; Jiang \& van den
Bosch 2015).  Second, whether or not the gamma-ray excess at
$\sim2$GeV from the Galactic Center (e.g., Calore \etal 2015) is
attributable to dark matter self-annihilation depends on the abundance
of subhaloes along the line-of-sight (e.g., Anderhalden \etal 2013;
Correa \etal 2015), which is subject to large uncertainties arising
from the halo-to-halo variance.  Third, the demographics of satellite
galaxies is directly related to that of subhaloes (e.g., Kravtsov
\etal 2004). As a consequence, models for the halo occupation
statistics of galaxies, which are used to interpret galaxy clustering
measurements, are guided by the occupation statistics of subhaloes. As
we will demonstrate, standard (oversimplified) assumptions regarding
the functional form of the halo-to-halo variance can result in
significant errors in the predicted two-point correlation functions of
satellite galaxies.  And finally, a proper treatment of the
halo-to-halo variance is of paramount importance when comparing
subhalo masses inferred from gravitational lensing with CDM
predictions (e.g., Vegetti \etal 2014; Okabe \etal 2014; Xu \etal
2015).

In this paper, we use a combination of cosmological $N$-body
simulations and a semi-analytical model of halo assembly and subhalo
evolution, to study the halo-to-halo variance of subhalo demographics
at unprecedented statistical power.  The semi-analytical model is
devised to quickly generate large ensembles of subhalo populations at
high mass resolution for a broad range of host halo mass and
cosmology.  It is described in detail in Jiang \& van den Bosch
(2016, Paper I), and its accuracy has been tested against multiple
simulations in van den Bosch \& Jiang (2016, Paper II). In particular,
as shown in Jiang \& van den Bosch (2015), the model accurately
reproduces the halo-to-halo variance in subhalo mass and velocity
functions, and is therefore ideally suited for this study, to
complement cosmological $N$-body simulations where they have limiting
statistical power. This happens both at the massive end, where the
abundance of host haloes is insufficient to accurately probe the
halo-to-halo variance, and at the low mass end, where the limiting
mass resolution only allows the study of subhalo statistics over a
tiny dynamic range in mass.

The paper is organized as follows.  In \S\ref{Sec:Method} we briefly
describe the semi-analytical model and the various $N$-body
simulations used. In \S\ref{Sec:MassFrac} we demonstrate that the
subhalo mass fraction is basically a function of halo formation time,
and we compare constraints on the subhalo mass fraction obtained using
gravitational lensing measurements to predictions from our model and
from a variety of numerical simulations. In \S\ref{Sec:sHOD} we focus
on the halo occupation distribution (HOD) of dark matter subhaloes,
$P(N|M)$, which expresses the probability that a host halo of mass $M$
contains $N$ subhaloes above a given mass limit. We demonstrate that
$P(N|M)$ deviates strongly, and in a complicated way, from a Poisson
distribution, and discuss the implications for HOD models of galaxy
clustering. We also address the dependence of the magnitude of the
non-Poissonity on halo formation redshift. Finally, in
\S\ref{Sec:MostMassiveSatellites}, we use the occupation statistics of
massive subhaloes to put constraints on the mass and formation time of
the Milky Way halo.  We summarize our findings in \S\ref{Sec:Summary}.

Throughout, we use lower- and upper-case letters to indicate subhalo
and host halo properties, respectively.  For example, $m$ and $\vmax$
represent the mass and maximum circular velocity of a subhalo, while
$M$ and $\Vmax$ correspond to the same quantities but for a host
halo. We use $\psi$ to either denote subhalo mass in units of the
present-day host halo virial mass, $m/M_0$, or the subhalo maximum
circular velocity in units of the present-day host halo virial
velocity, $\vmax/\Vvir$.  Throughout, we define halo formation time as
the earliest redshift ($\zf$) by which a halo has assembled half of
its present-day mass.


\section{Method}
\label{Sec:Method}


\subsection{$N$-body Simulations}  
\label{Sec:Sim}

The main simulations that we use in this study are the Bolshoi (Klypin
\etal 2011) and MultiDark (Prada \etal 2012) simulations.  Both follow
the evolution of $2048^3$ dark matter particles using the Adaptive
Refinement Tree (\ART) code (Kravtsov, Klypin \& Khokhlov 1997) in a
flat $\Lambda$CDM cosmology with parameters
$(\Omega_{\rmm,0},\Omega_{\Lambda,0}, \Omega_{\rmb,0},h,\sigma_8,
n_\rms)=(0.27,0.73,0.047,0.7,0.82,0.95)$.  Bolshoi has a particle mass
of $m_\rmp = 1.35\times 10^8 \Msunh$ and a box of size $250\mpch$,
while MultiDark has a particle mass of $m_\rmp = 8.7\times 10^9
\Msunh$ and a box of size $1000\mpch$.  As shown in Paper II, due to
the limited mass resolution, the evolved subhalo mass functions become
incomplete below 50 particles, which corresponds to a mass scale of
$\mres = 10^{9.83}\Msunh$ and $10^{11.64}\Msunh$ for Bolshoi and
MultiDark, respectively.

We also use published results or publicly available data from several
suites of zoom-in simulations, including Rhapsody (Wu \etal 2013),
ELVIS (Garrison-Kimmel \etal 2014a), Phoenix (Gao \etal 2012), Aquarius
(Springel \etal 2008), and COCO (Bose \etal 2016).  These are obtained
by re-simulating specific haloes or sub-volumes, extracted from
medium-resolution cosmological volumes, at much higher resolution.
Details of all simulations used throughout this work are summarized in
Table \ref{tab:sims}, and we refer interested readers to the original
papers for details.

Haloes and subhaloes in Bolshoi, MultiDark, Rhapsody, and ELVIS are
identified using the \Rockstar halo finder (Behroozi \etal2013a,b),
while COCO, Phoenix and Aquarius results are obtained using \Subfind
(Springel \etal 2001).  \Rockstar is a phase-space halo finder, which
uses adaptive, hierarchical refinement of friends-of-friends groups in
six phase-space dimensions and one time dimension. \Subfind, on the
other hand, only relies on information in configuration-space. As
discussed in Paper II, \Subfind has a tendency to underestimate the
masses of subhaloes close to the center of the host halo, which has an
appreciable impact on the massive end of the subhalo mass function.

Halo mass is defined as the mass enclosed in a sphere with an average
density of $\Delta$ times the critical density of the Universe. For
Bolshoi, MultiDark, Rhapsody, and ELVIS $\Delta = \Delta_{\rm vir}
\sim 100$, with $\Delta_{\rm vir}$ given by the fitting function of
Bryan \& Norman (1998), while the halo catalogs of COCO, Phoenix, and
Aquarius are based on $\Delta = 200$.  When needed, we convert
$M_{200}$ to $M_{\rm vir}$ using the average concentration mass
relation of Macc\`io et al. (2008).
\begin{table*}\label{tab:sims}
\caption{Numerical Simulations used in this Paper}
\begin{center}
\tabcolsep=0.05cm
\begin{tabular}{lcccccccccccl}
\hline\hline
 Simulation & $\Omega_{\rmm,0}$ & $\Omega_{\Lambda,0}$ & $\Omega_{\rmb,0}$ & $\sigma_8$ & $n_\rms$ & $h$ & $L_{\rm box}$ & $N_\rmp$ & $m_\rmp$ & $N_{\rm halo}$ & $M_{\rm halo}$ & Reference \\
  &  &  &  & &  & & $\mpch$ &  & $\Msunh$ &  & $\Msunh$ &  \\
\hline
Bolshoi      & 0.27 & 0.73 & 0.047 & 0.82 & 0.95 & 0.70 &   250 & $2048^3$ & $1.35 \times 10^8$ &  -- & -- & Klypin \etal (2011) \\ 
MultiDark      & 0.27 & 0.73 & 0.047 & 0.82 & 0.95 & 0.70 &   1000 & $2048^3$ & $8.7 \times10^9$ & -- & -- & Prada \etal (2012) \\ 
\hline
Rhapsody    & 0.25 & 0.75 & 0.04 & 0.8 & 1.0 & 0.70 & -- & $8192^3$ & $1.3 \times 10^8$ & 96 & $\Mvir=10^{14.8\pm0.05}$ & Wu \etal (2013) \\ 
ELVIS    & 0.266 & 0.734 & 0.045 & 0.801 & 0.963 & 0.70 & -- & -- & $1.35 \times 10^5$ & 48 & $\Mvir=10^{12.08\pm0.23}$  & Garrison-Kimmel \etal (2014a) \\
COCO   & 0.272 & 0.728 & 0.045 & 0.81 & 0.967 & 0.704 & 17.4 & $1.3\times10^{10}$ & $1.14 \times 10^5$ & -- & -- & Bose \etal (2016) \\
Phoenix   & 0.25 & 0.75 & 0.045 & 0.9 & 1.0 & 0.73 & -- & -- & $6.4 \times 10^5$ & 9 & $M_{200}=10^{15.0\pm0.3}$ & Gao \etal (2012) \\
Aquarius   & 0.25 & 0.75 & 0.045 & 0.9 & 1.0 & 0.73 & -- & -- & $1.7 \times 10^3$ & 6 & $M_{200}=10^{11.93\pm0.18}$ & Springel \etal (2008) \\
\hline\hline
\end{tabular}
\end{center}
\medskip
\begin{minipage}{\hdsize}
 Note: COCO re-simulates a volume which approximates a sphere of radius
$17.4\mpch$ at the present time, extracted from a parent volume of
$70.4\mpch$. The other zoom-in simulations re-simulate
individual haloes rather than a coherent sub-volume.  We only focus on
the COCO result for haloes with $M_{50}=10^{12.25\pm0.25}\Msunh$.
\end{minipage}
\end{table*}
%


\subsection{Semi-analytical model}  
\label{Sec:Model}

In addition to the $N$-body simulations described above, we also use
the semi-analytical model of substructure evolution developed in Jiang
\& van den Bosch (2016). Here we briefly describe the model, and refer
interested readers to Paper I for more details.  The model constructs
halo merger trees using the algorithm developed by Parkinson \etal
(2008), uses simple semi-analytical descriptions to evolve subhalo
mass and structure, considers the entire hierarchy of substructure,
and includes an empirical treatment of subhalo disruption.  As shown
in Jiang \& van den Bosch (2014), the merger statistics from the
Parkinson \etal algorithm are in excellent agreement with those from
numerical simulations. Throughout this paper, we adopt the Bolshoi
cosmology for the model, and we have verified that changing the
cosmological parameters to any of the other cosmologies listed in
Table 1 only results in very small differences (cf., Paper II and
Dooley \etal 2014).

The mass of a subhalo is evolved using an orbit-averaged mass loss
rate, $\dot{m}=f[m(t),M(t)]$, which takes a functional form that is
motivated by a simple toy model of tidal stripping and has two free
parameters: one determines the amplitude and the other controls the
$m/M$-dependence.  For each subhalo, the amplitude is drawn from a
log-normal distribution that accounts for the scatter in orbital
energies, orbital angular momenta and orbital phases.

The maximum circular velocity, $\vmax$, and the corresponding radius,
$\rmax$, of a subhalo are computed using the empirical relations of
Pe\~narrubia \etal (2010), $\vmax=g_1(\vacc, m/\macc)$ and
$\rmax=g_2(\racc, m/\macc)$.  Note that $\vacc$ and $\racc$ represent
the $\vmax$ and $\rmax$ of a subhalo {\it at accretion}, which are
computed by assuming an Navarro, Frenk \& White (1997, NFW hereafter) density profile with a concentration
$c(\macc,\tacc)$. The latter is determined from the mass assembly
history of the subhalo progenitor prior to accretion using the model
of Zhao \etal (2009).  As such, the scatter in the merger histories of
subhalo progenitors is imprinted in the structure of the evolved
subhaloes.  It turns out that the model prediction for the joint
distribution of $\vmax$ and $\rmax$ of evolved subhaloes is
indistinguishable from that in high-resolution simulations (see Jiang
\& van den Bosch 2015).

Following Taylor \& Babul (2004), we consider a subhalo disrupted once
its mass, $m(t)$, drops below a critical value $\macc(<\fdis\rsacc)$,
i.e., the mass enclosed at accretion within a radius that is $\fdis$
times the NFW scale radius at accretion ($\rsacc$).  Like the mass
loss rate, the critical mass also varies from one subhalo to another:
for each subhalo, we draw a value for $\fdis$ from a log-normal
distribution that approximates the $\fdis$ distribution of disrupting
subhaloes in the Bolshoi simulation. Subhalo disruption is an
indispensable ingredient of the model. As discussed in Paper I, one
can construct a model without disruption that perfectly fits the
evolved subhalo mass function (by enhancing the mass loss rates), but
such a model predicts retained mass fractions, $m/\macc$, that are
much lower than what is found in the simulation.

The advantage of the model is that it is extremely fast, which makes
it useful to complement numerical simulations where they have limiting
statistical power. For example, the Bolshoi simulation resolves the
substructure in cluster-sized haloes ($M_0 \simeq
10^{14.5-15.0}\Msunh$) all the way down to $\sim 10^{-4.5}
M_0$. However, because the Bolshoi simulation only covers $\sim 0.16
h^{-3} \Gpc^3$, the actual number of cluster-sized haloes is too small
for a reliable statistical analysis.  The larger volume of the
MultiDark simulation results in a much larger number of massive host
haloes, but their subhaloes can only be resolved down to $\sim 10^{-3}
M_0$. With the semi-analytical model, on the other hand, it is trivial
to simulate thousands of massive host haloes with a mass resolution of
$10^{-5} M_0$ in a matter of hours. As an example, Jiang \& van den
Bosch (2015) used the semi-analytical model to simulate tens of
thousands of Milky-Way size host haloes with a mass resolution of
$10^{-5} M_0$. For comparison, although the Bolshoi simulation
contains hundreds of thousands of such Milky-way sized host haloes, it
only resolves their substructure down to $\sim 10^{-2} M_0$.
\begin{figure}
\centerline{\psfig{figure=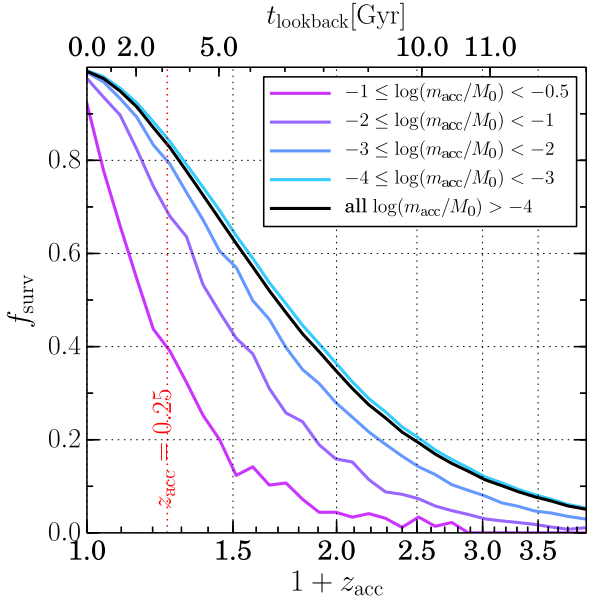,width=0.5\hdsize}}
\caption{The surviving fraction of subhaloes as function of accretion
  redshift, $\zacc$, for subhaloes with different masses at accretion,
  $\macc$, in host haloes with $M_0=10^{13.5}\Msunh$ (we have verified
  that the results show little dependence on host halo mass).
  Coloured lines indicate different bins in $\log(\macc/M_0)$, as
  indicated, while the black line represents all subhaloes with
  $\macc>10^{-4}M_0$. The vertical, red dotted line at $\zacc=0.25$
  indicates the average accretion redshift for subhaloes that are
  currently at their first apo-center.}
\label{Fig:Survivability}
\end{figure}
%


\subsubsection{The Prevalence of Subhalo Disruption}
\label{Sec:Disruption}

Tidal stripping and impulsive encounters with the host halo and with
other subhaloes may ultimately result in the complete disruption of a
subhalo. Subhalo disruption is prevalent in numerical simulations, to
the extent that roughly half of all subhaloes ever accreted has been
disrupted by $z=0$ (e.g., Han \etal 2016; Paper I). To make this more
quantitative, we use the semi-analytical model described in
\S\ref{Sec:Model} above to construct 5000 model realizations of a host
halo with mass $M_0=10^{13.5}\Msunh$\footnote{Results for other host
  halo masses are very similar}. We compute the fraction of subhaloes
that survive to the present day as a function of their mass and
redshift at accretion. Results are shown in
Fig.~\ref{Fig:Survivability}, where the thick, black line indicates
the surviving fraction of all subhaloes with $\macc/M_0 > 10^{-4}$ as
a function of their accretion redshift. As is evident, less than 40\%
(10\%) of subhaloes accreted at $\zacc=1$ (2) survive to the present
day. The colored lines indicate the surviving fractions among
subhaloes in different bins of $\macc/M_0$ (as indicated). There is a
clear tendency for more massive subhaloes to undergo more efficient
disruption. As many as 80\% (95\%) of all subhaloes with $\macc >
0.01M_0$ ($\macc > 0.1M_0$) have been disrupted since $\zacc = 1$.
The average accretion redshift of subhaloes which have just completed
their first orbit, and are currently at their apo-centers, is $\zacc
\sim 0.25$ (van den Bosch \etal 2016).  Based on this, we estimate
from Fig.\ref{Fig:Survivability} that $\sim20\%$ of subhaloes with
$\macc > 10^{-4} M_0$ have been disrupted during their first orbital
period. This fraction increases to $30\%$ and $60\%$ for subhaloes
with $0.01\le \macc < 0.1M_0$, and $\macc\ge0.1M_0$, respectively.

We caution that the efficiency of subhalo disruption depicted in
Fig.\ref{Fig:Survivability} is based on the model, which is tuned to
reproduce subhalo evolution in the Bolshoi simulation.  It is not yet
clear to what extent subhalo disruption is due to numerical artifacts
as opposed to the physical impact of tidal stripping and
heating. Recall that it wasn't until the end of the 1990's that
numerical simulations started to reach sufficient mass and force
resolution to resolve a surviving population of subhaloes (e.g.,
Moore, Katz \& Lake 1996; Tormen, Bouchet \& White 1997; Ghigna \etal
1998; Klypin \etal 1999). And even today, the limiting mass and force
resolution of numerical simulations may well result in a continued
overmerging of substructure, especially near the centers of their host
haloes.  We will address the issue of numerical disruption in more
detail in a separate study (van den Bosch, in prep.). For now, it is
important to be aware that the results presented here are only as
accurate as the simulations used.
\begin{figure*}
\centerline{\psfig{figure=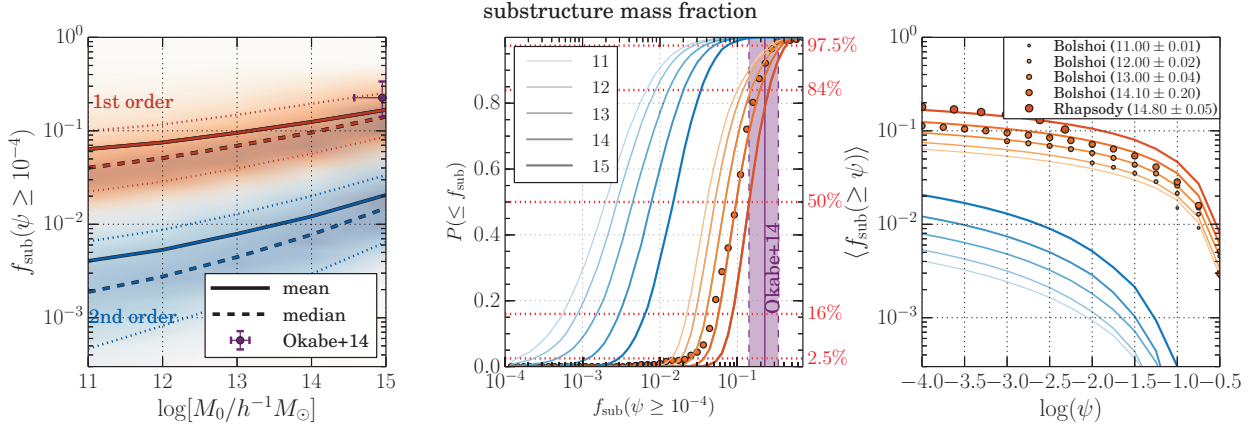,width=1.0\hdsize}}
\caption{{\it Left-hand panel}: Mass fraction in present-day,
  surviving first-order (orange) and second-order (blue) subhaloes
  with $m\ge10^{-4}M_0$, as a function of host halo mass.  The solid,
  dashed and dotted lines represent the average, median, and the 16th
  and 84th percentiles respectively.  The data point with errorbars is
  the weak lensing measurement of the subhalo mass fraction in the
  Coma cluster from Okabe \etal (2014).  {\it Middle panel}: the
  cumulative distribution of subhalo mass fraction for five different
  host halo masses, as indicated.  The vertical band indicates the
  Okabe \etal (2014) measurement.  {\it Right-hand panel}: the average
  subhalo mass fraction, $\langle \fsub (\ge \psi) \rangle$ as a
  function of the mass threshold $\psi = m/M_0$.}
\label{Fig:MassFrac}
\end{figure*}
%


\section{Halo-to-Halo Variance of the Subhalo Mass Fraction}
\label{Sec:MassFrac}

We now use our model and the simulations listed in Table~1 to
investigate the halo-to-halo variance in the subhalo mass fraction,
which we define as
\begin{equation}\label{Eq:DefineMassFrac}
\fsub (\ge\psi) = \frac{1}{M_0}\sum\limits_{i} m_{i}
\Theta(\psi_{i}-\psi) = \int_{\psi}^{1} {\rmd N \over
  \rmd\ln\psi^\prime} \rmd \psi^\prime\,.
\end{equation}
Here $\sum_{i}$ indicates summation over all subhaloes, $\Theta(x)$
is the Heaviside step function, $\psi$ is shorthand notation for
$m/M_0$, and $\rmd N/\rmd\ln\psi$ is the subhalo mass function.
Throughout this section, we consider a lower limit of $\psi=10^{-4}$
[i.e., $\fsub = \fsub(\ge10^{-4})$], unless stated otherwise.

As discussed in detail in Paper~I, the subhalo mass fraction is the
outcome of the competition between the accretion of new subhaloes and
the dynamical evolution (tidal stripping and disruption) of existing
subhaloes.  At $z \sim 0$, the mass-loss timescale of subhaloes is of
the order of the dynamical time ($\sim 2$Gyr) and scales with redshift
as $(1+z)^{-3/2}$ (e.g., Paper I). The time scale for the accretion of
subhaloes is equivalent to the time scale of halo mass assembly, and
is of the order of $M/\dot{M} \sim 10$Gyr, with a similar redshift
scaling at low $z$ (e.g., McBride \etal 2009).  Hence, at the present
day most host haloes are in the regime of net subhalo mass loss (i.e.,
$\fsub$ decreases with time). As a consequence, the subhalo mass
fraction is strongly (anti-)correlated with the halo formation time
(see Paper~I), which is why it has been used as an indicator of the
level of relaxedness of a halo (e.g., De Lucia \etal 2004; Shaw \etal
2006; Ludlow \etal 2013).  In the context of flux-ratio anomalies of
multiply-imaged quasars, the frequency and strength of anomalies scale
with $\fsub$ (Dalal \& Kochanek 2002).  Hence, a meaningful
interpretation of flux-ratio anomalies requires a detailed
understanding of the expected distribution of subhalo mass fractions.

In what follows, we first characterize the $\fsub$ distribution as a
function of host halo mass (\S\ref{Sec:MassFracHaloMassDependence}),
and then show that the halo mass dependence of the average mass
fraction $\langle \fsub \rangle$ essentially reflects the fact that
more massive haloes form later (\S\ref{Sec:MassFraczformDependence}).
We consider both subhaloes and the subhaloes of subhaloes, which we
refer to as first-order and second-order substructure, respectively.
Finally, we compare the model predictions for $\fsub$ with constraints
inferred from gravitational lensing (\S\ref{Sec:MassFracRadialProfile}).


\subsection{Halo Mass Dependence} \label{Sec:MassFracHaloMassDependence} 

The left-hand panel of Fig.\ref{Fig:MassFrac} plots the subhalo mass
fraction as a function of host halo mass.  On average, $\fsub$
increases with host halo mass, from $\sim6\%$ at $M_0=10^{11}\Msunh$
to $\sim17\%$ at $M_0=10^{15}\Msunh$.  The relation can be
approximated by
\begin{equation} \label{Eq:fsubM0}
\log\langle\fsub\rangle = a + b\log\left(\frac{M_0}{10^{12}\Msunh}\right),
\end{equation}
with $(a,b)=(-1.12, 0.12)$ and $(-1.28, 0.13)$ good descriptions of
the mean and median relations, respectively.  For second-order
subhaloes, the corresponding best-fit parameters are
$(a,b)=(-2.26,0.17)$ and $(-2.53,0.22)$, respectively. Hence, the mass
fraction of second-order subhaloes is roughly 10\% of that of
first-order subhaloes, which themselves make up roughly 10\% of the
mass of the host halo. This self-similarity is simply a manifestation
of the self-similar nature of structure formation (cf. van den Bosch
\etal 2014).  Note that our halo mass definition is `inclusive', i.e.,
the mass of a first-order subhalo includes the mass of all its
second-order subhaloes.

There is a weak trend for the scatter in $\fsub$ to be smaller for
more massive haloes. In addition, for a given halo mass, the
second-order $\fsub$ has larger scatter than that of first-order
subhaloes.  We find that the $\fsub$ distributions can be approximated
by log-normal distributions with the mean $\log\langle\fsub\rangle$
given by Eq.(\ref{Eq:fsubM0}), and the standard deviation given by
\begin{equation} \label{Eq:sigfsubM0}
\sigma_{\log\fsub} = c + d \log\left(\frac{M_0}{10^{12}\Msunh}\right),
\end{equation}
with $(c,d)=(0.26, -0.03)$ and $(0.5,-0.03)$ for first-order and
second-order subhaloes, respectively.

The middle panel of Fig.\ref{Fig:MassFrac} plots the distributions of
$\fsub$ for host haloes with mass
$M_0=10^{11},10^{12},...,10^{15}\Msunh$ (solid lines), based on 10,000
model realizations for each halo mass.  In order to show that the
model predictions are similar to the results from $N$-body
simulations, we compare the $\fsub$ distribution for model
realizations of haloes with $M_0=10^{14}\Msunh$ to that for the 372
Bolshoi haloes with $M_0=10^{14.1\pm0.2}\Msunh$.  This mass bin is
chosen as a compromise between the need for a sufficient number of
host haloes and the requirement that subhaloes are complete down to
$m>10^{-4}M_0$.  The simulation results are in excellent agreement
with the model predictions, although the former reveals a slightly
more extended low-$\fsub$ tail.

The right-hand panel of Fig.\ref{Fig:MassFrac} plots the average mass
fraction, $\langle \fsub (\ge \psi) \rangle$, as a function of
threshold mass, $\psi$.  From $\psi=10^{-2}$ to $10^{-4}$, the average
mass fraction in first-order (second-order) subhaloes increases by a
factor of $\sim2$ ($\sim4$).  We compare the average, first-order
subhalo mass fraction, $\langle \fsub (\ge \psi) \rangle$, with that
from numerical simulations.  As before, the model predictions are
based on 10,000 realizations of haloes with $M_0 =
10^{11},10^{12},...,10^{15} \Msunh$.  The simulation results are based
on the 16969, 4786, 1138, and 372 haloes with$M_0 =
10^{11\pm0.01},10^{12\pm0.02},10^{13\pm0.04}$, and
$10^{14.1\pm0.2}\Msunh$ from the Bolshoi simulation, and the 96 haloes
with $M_0 = 10^{14.8\pm0.05}\Msunh$ in the Rhapsody zoom-in suites,
respectively.  Good agreement is achieved for all the mass scales.

In a recent study, Okabe \etal (2014) used weak gravitational lensing
to measure the mass of subhaloes in the Coma cluster. Adopting a host
halo mass of $M_0 = 8.92^{+20.05}_{-5.17} \times 10^{14}\Msunh$ (Okabe
\etal 2010) they infer a subhalo mass fraction $\fsub(>10^{-3}) =
0.226^{+0.111}_{-0.085}$. This measurement is indicated as a purple
data point in the left-hand panel and as a purple vertical band in the
middle panel of Fig.\ref{Fig:MassFrac}\footnote{Note that the model
  predictions are for $\fsub(>10^{-4}$); however, as is evident from
  the right-hand panel of Fig.\ref{Fig:MassFrac}, the subhalo mass
  fractions $\fsub(>10^{-3})$ and $\fsub(>10^{-4})$ are very similar
  for massive host haloes.}. As is evident, the Okabe \etal result
is significantly higher than the average mass fraction predicted by
our model, or by the numerical simulations (i.e., according to our
model $\langle \fsub(>10^{-3}) \rangle = 0.145^{+0.026}_{-0.017}$).
However, since the halo-to-halo variance is large, we find that the
Okabe \etal measurement falls within the 95 percentile of the
distribution, and can therefore be considered in agreement with
$\Lambda$CDM predictions.
\begin{figure*}
\centerline{\psfig{figure=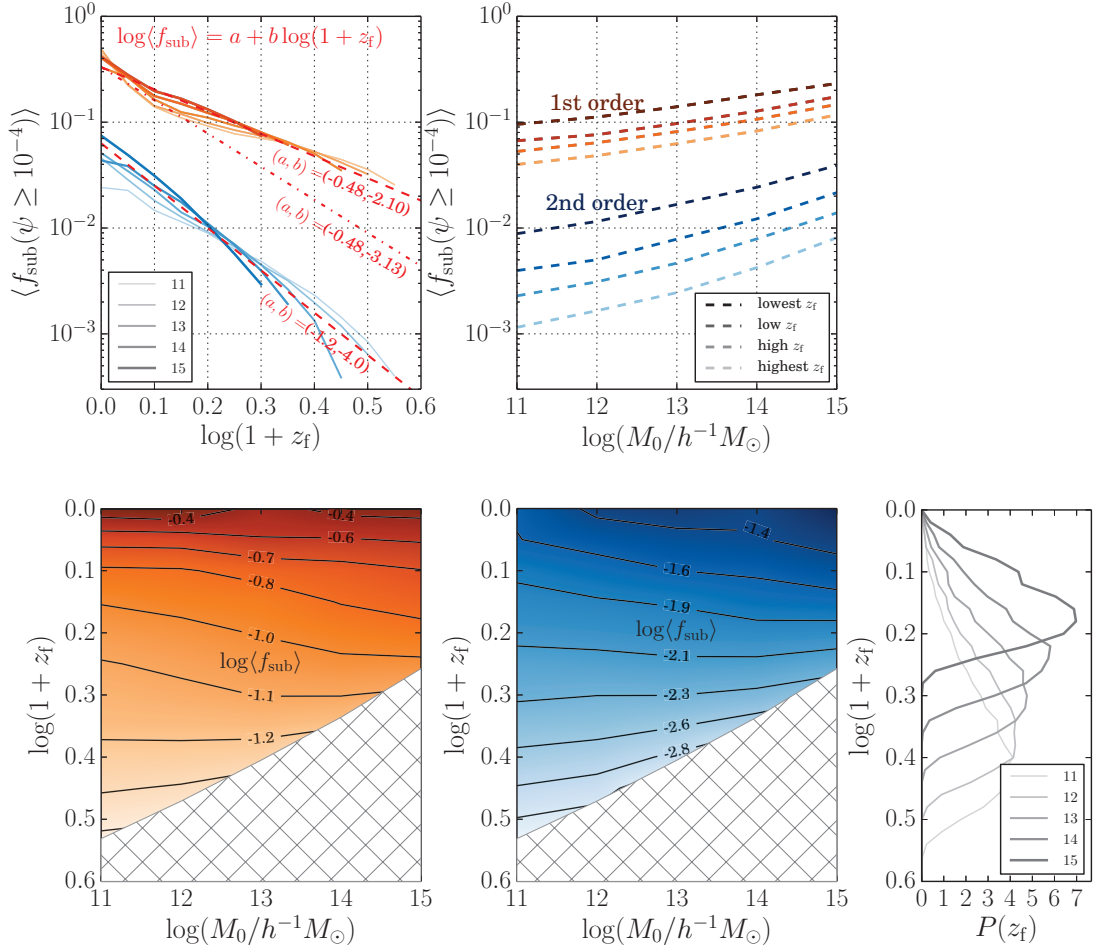,width=0.85\hdsize}}
\caption{{\it Upper, left-hand panel}: the average mass fraction in
  subhaloes with $m>10^{-4}M_0$, as a function of formation redshift,
  for different halo masses, as indicated.  The red dashed lines are
  fitting functions of the form of Eq.~(\ref{Eq:fsubzf}) with
  parameters indicated in the panel. For comparison, the dash-dotted
  line represents the Giocoli \etal (2010) fitting function for
  first-order subhaloes.  {\it Upper, right-hand panel}: $\langle
  \fsub \rangle$ as a function of $M_0$ for haloes of different $\zf$.
  Lines of different darkness correspond to different quartiles of the
  $\zf$-distribution, as indicated.  {\it Lower panels}: contours of
  $\langle \fsub \rangle$ in the $\zf$-$M_0$ space of first-order
  (left) and second-order (right) subhaloes.  {\it Side panel}: $\zf$
  distributions of haloes of different masses, as indicated.}
\label{Fig:MassFraczformDependence}
\end{figure*}
%


\subsection{Formation Redshift Dependence} \label{Sec:MassFraczformDependence}

The upper, left-hand panel of Fig.\ref{Fig:MassFraczformDependence}
plots the average subhalo mass fraction as a function of formation
time, for $M_0=10^{11},10^{12},...,10^{15}\Msunh$.  Earlier-forming
haloes have lower subhalo mass fractions, as their subhaloes were
accreted earlier and thus exposed to tidal evolution for a longer
period of time (e.g., Zentner \etal 2005; van den Bosch \etal 2005;
Paper I).  The relation between $\langle \fsub \rangle$ and $\zf$
exhibits little dependence on halo mass, and can be approximated by
\begin{equation} \label{Eq:fsubzf}
\log\langle\fsub\rangle = a + b\log(1+\zf),
\end{equation}
with $(a,b)=(-0.48,-2.10)$ and $(-1.2,-4.0)$ for first- and
second-order subhaloes, respectively.

Giocoli \etal (2010a) find a $\langle \fsub \rangle$-$\zf$ relation of
the functional form of Eq.(\ref{Eq:fsubzf}) but with
$(a,b)=(-3.13,-0.48)$, using the GIF2 simulation.  This seemingly
contradicts our results: for halos with $\zf \sim 1$ the subhalo mass
fraction is a factor of $\sim 2$ lower than our model predictions.
However, their subhaloes are selected using an absolute mass cut that
corresponds to $\psi \sim 10^{-2}$ for the bulk of their sample of
host haloes.  Since $\langle \fsub(>10^{-2}) \rangle$ is a factor of
$\sim 2$ lower than $\langle \fsub(>10^{-4})\rangle$, as can be seen
from the right-hand panel of Fig.\ref{Fig:MassFrac}, the difference
between the results of Giocoli \etal and this work simply reflects a
mass-resolution effect.

The upper, right-hand panel of Fig.\ref{Fig:MassFraczformDependence}
plots the average subhalo mass fraction as a function of halo mass.
Different lines correspond to different quartiles in the distribution
of $\zf$, as indicated. As is evident, the youngest quartile has a
factor of $\sim 2$ (5) more mass in first-order (second-order)
subhaloes than the oldest quartile.

Finally, the lower panels of Fig.\ref{Fig:MassFraczformDependence}
show contours of $\langle \fsub \rangle$ in the formation
redshift-halo mass space.  We find that, for a given $\zf$, $\langle
\fsub \rangle$ is almost independent of $M_0$.  The hatched areas mark
the regions in parameter space that contain very few haloes, as can be
seen from the side-panel, which shows the $\zf$ distributions for
different halo masses. As is evident, to a good approximation $\langle\fsub\rangle$
is just a function of $\zf$, while the mass dependence simply reflects
that more massive haloes assemble later (on average). We emphasize,
though, that a halo's formation time does not completely determine its
substructure content, as even for a fixed formation time, the subhalo
mass fraction reveals significant halo-to-halo variance.


\subsection{Radial Profile} \label{Sec:MassFracRadialProfile}

Dark matter subhaloes can give rise to flux-ratio anomalies in
multiply imaged quasars (Dalal \& Kochanek 2002; Nierenberg \etal
2014) and can cause surface brightness perturbations in the images of
Einstein rings and lensed arcs (e.g., Koopmans 2005; Vegetti \&
Koopmans 2009; Vegetti \etal 2010, 2012, 2014; Hezaveh \etal 2016).
Hence, strong gravitational lensing provides a unique window on the
substructure content of dark matter haloes.

In particular, Vegetti \etal (2014) searched for subhaloes in a sample
of 11 early-type lens galaxies with stellar masses of
$\simeq10^{11.5}\Msunh$ (Auger \etal 2009) at $z\simeq0.2$ using the
Sloan Lens ACS Survey (SLACS, Bolton \etal 2006).  They detected
subhaloes only in two systems, one with a mass $m \simeq 3.5 \times
10^9\Msun$ at an Einstein radius of $R_\rmE \simeq 4.6\kpc$, the other
with a mass $m \simeq 1.7 \times 10^{10}\Msun$ at $R_\rmE \simeq
5.05\kpc$.  Combining these two detections with the non-detections,
they infer a projected subhalo mass fraction of $\Fsub =
0.0064^{+0.0080}_{-0.0042}$ (68\% CL) within a median Einstein radius
of $\langle R_\rmE \rangle \simeq 4.2\kpc$.

The subhalo mass detection threshold of this SLACS sample is of the
order of $10^8 \Msunh$ (Vegetti \etal 2010), while the characteristic
host halo mass is $10^{13.1} \Msunh$. We obtain the latter by applying
the Moster, Naab, \& White (2013) stellar-to-halo mass relation (at
$z=0.2$) to the stellar masses published in Vegetti \etal
(2014). Hence, a fair comparison with $\Lambda$CDM predictions
requires simulations to resolve subhaloes down to $m \simeq10^{-5}
M_0$, for which it is challenging to obtain a statistically large
sample of haloes.  The ELVIS haloes and the cluster-size haloes in the
Bolshoi simulation have sufficient resolution, but constitute a small
sample of no more than 100 haloes.  In addition, their masses differ
from the characteristic host halo mass of the SLACS sample.  On the
other hand, our semi-analytical model can generate large samples of
$10^{13} \Msunh$ host haloes at sufficiently high mass resolution, but
provides no information regarding the radial distribution of subhaloes
within their hosts. As a compromise, we use the simulations to extract
information regarding the spatial distribution of subhaloes, which we
apply to a statistically large sample of $10^{13.1} \Msunh$ host haloes
generated with the model. This allows us to predict the halo-to-halo
variance of the projected subhalo mass fraction profile, thereby
allowing for a fair comparison with the lensing data.
\begin{figure*}
\centerline{\psfig{figure=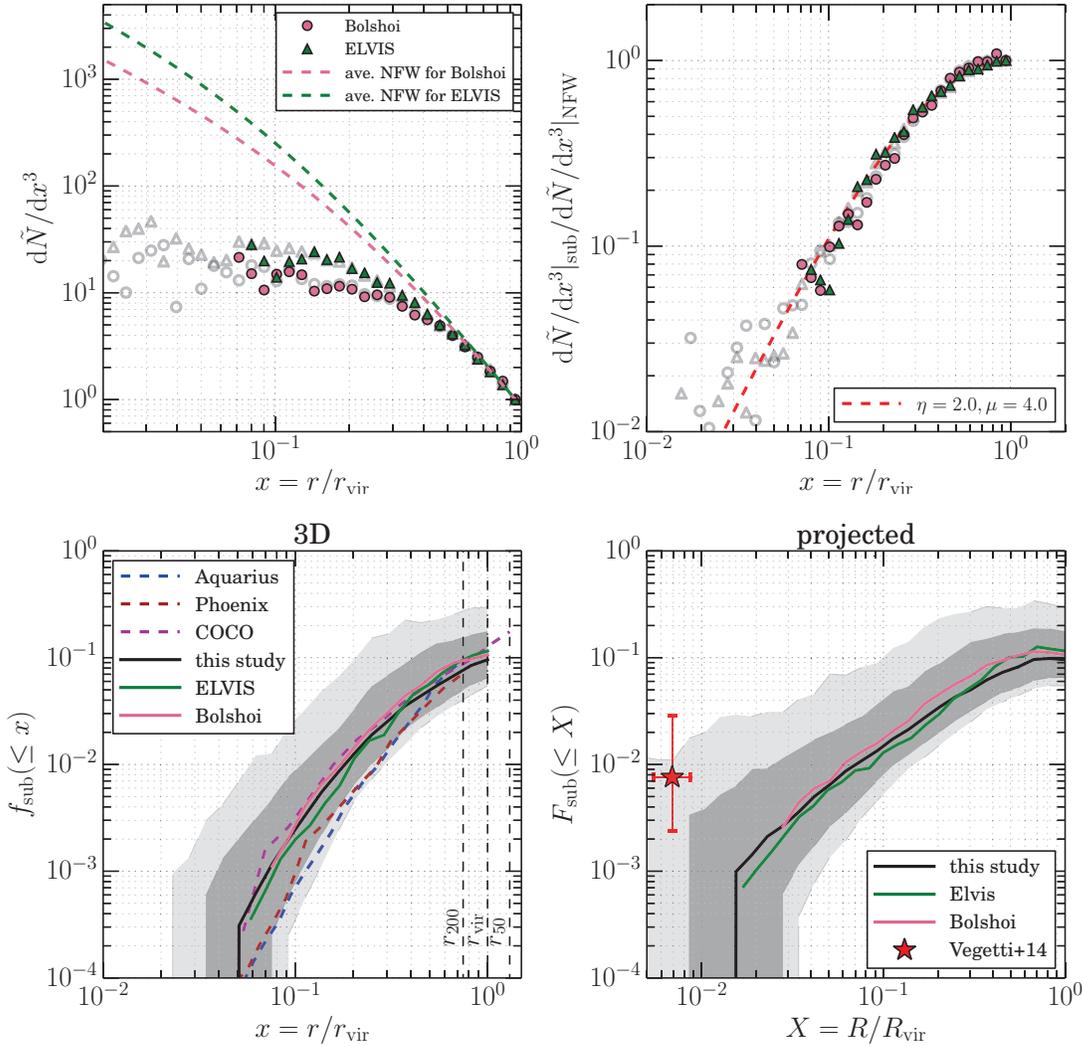,width=0.85\hdsize}}
\caption{{\it Upper left-hand panel}: radial subhalo number
  distributions for the 48 ELVIS haloes (triangles) and the 38 Bolshoi
  haloes (circles) with $M_0>10^{14.5}\Msunh$), normalized at $\rvir$.
  Filled and open symbols indicate the median and average,
  respectively.  The dashed lines are the average NFW profiles for the
  host haloes. {\it Upper right-hand panel:} the `bias function',
  defined as the ratio of the normalized subhalo number distribution
  to the normalized density profile of the corresponding host haloes.
  The `bias' functions for the different simulation results are
  indistinguishable and well fit by Eq.~(\ref{Eq:BiasFunction}) with
  $\eta=2.0$ and $\mu=4.0$, as indicated by the red, dashed line. {\it
    Lower left-hand panel:} the median subhalo mass fraction within
  radius $x=r/\rvir$, in 3D, for different simulations (coloured
  lines, as indicated) and for the model (solid, black line).  Note
  that the model prediction is obtained by drawing subhalo positions
  according to the universal `bias function', and by assuming no
  radial mass segregation of subhaloes. {\it Lower right-hand panel:}
  same as lower left-hand panel, but in 2D (for a randomly chosen
  line-of-sight).  The data point with error bars is the gravitational
  lensing measurement from Vegetti \etal (2014). In both lower panels
  the darker and lighter grey bands indicate the 68\% and 90\%
  confidence intervals around the median.}
\label{Fig:Fsub}
\end{figure*}

We start by computing the radial distribution of subhaloes using the
48 Milky Way-size haloes with $M_0 = 10^{12.08 \pm 0.23}$ in the ELVIS
simulation and the 38 cluster-size haloes with $M_0 > 10^{14.5}\Msunh$
in the Bolshoi simulation. In both cases we include all subhaloes with
$m>10^{-5}M_0$. The upper left-hand panel of Fig.\ref{Fig:Fsub} plots
the average number of subhaloes per unit shell volume as a function of
the halo-centric distance in units of the virial radius, $r/\rvir$,
and normalized to unity at $\rvir$. As is evident, the radial
distribution of subhaloes in the cluster-size host haloes from the
Bolshoi simulation is less concentrated than that of the Milky-Way
size host haloes in ELVIS.  The dashed lines correspond to the average
dark matter density profiles of the corresponding host haloes, equally
normalized at the virial radii. They show that the host haloes in
Bolshoi are also less concentrated on average ($\langle \cvir \rangle
\simeq 5.6$) than those in the ELVIS simulations ($\langle \cvir
\rangle \simeq 9.8$). The upper right-hand panel of
Fig.~\ref{Fig:Fsub} plots the average `radial bias' functions, defined
as the ratio between the average (normalized) subhalo distribution and
the average (normalized) NFW density profile of their host
haloes. Note how the Bolshoi and ELVIS results show radial bias
functions that are in excellent mutual agreement. They are well fit by
\begin{equation}\label{Eq:BiasFunction}
\phi(x) \equiv 
\frac{\rmd \tilde{N}/\rmd x^3|_{\rm sub}}{\rmd \tilde{N}/\rmd x^3|_{\rm NFW}}
= 2^\mu \frac{x^{\eta}}{(1+x)^{\mu}}, \quad x\equiv r/\rvir
\end{equation}
with $(\eta,\mu)=(4.0,2.0)$, shown as the red, dashed line. Hence, in
what follows we assume that the radial bias function, $\phi(x)$, is
given by this fitting function, with no dependence on halo mass
\footnote{We note this is in conflict with Han \etal (2016), who
  expressed the bias function as $\phi(x)=x^\gamma$, and found $\gamma
  = 1.33$ and $\gamma = 0.95$ for Milky Way-size and cluster-size
  haloes, respectively. However, their results are based on a
  different halo finder, and on different definitions for halo mass
  and radius.} We also assume that the radial bias function is
independent of the mass of the subhaloes. This is a valid assumption,
since van den Bosch \etal (2016) have shown that there is little to no
radial segregation of subhaloes by {\it present-day} mass.

Using the best-fit radial bias function of Eq.(\ref{Eq:BiasFunction}),
we assign halo-centric radii to subhaloes generated by our
semi-analytical model, by randomly drawing unitless halo-centric
radii, $x=r/\rvir$ from the probability distribution $P(x) = 4 \pi \,
x^2 \, \phi(x) \, \rho(x)$, where $\rho(x)$ is the NFW density profile
of the host halo in question. Next we compute the subhalo mass
fraction profile as
\begin{equation}\label{Eq:fsub}
\fsub(<x) = \left[ \sum\limits_{x_\rmi<x} m_\rmi \right] / M(<x|\cvir,M_0),
\end{equation}
where the halo mass within radius $x$ is given by $M(<x) = M_0 \times
f(\cvir x)/f(\cvir)$, with $f(x)=\ln(1+x)-x/(1+x)$.

In order to obtain a {\it projected} mass fraction profile, we assume
that subhaloes are distributed isotropically with respect to their
host centers and randomly choose a line-of-sight, so that the
projected radii can be computed as $X = x\sin(\theta)$, where $\theta
= \arccos(2\calR-1)$ with $\calR$ a random number drawn from a uniform
distribution over $[0,1)$.  The projected subhalo mass fraction
  profile is computed as
\begin{equation}\label{Eq:Fsub}
\Fsub(<X) = \left[\sum\limits_{X_\rmi<X} m_\rmi \right] / M(<X|\cvir,M_0),
\end{equation}
where the halo mass within the projected radius $X$ is given by $M(<X)
= M_0 \times g(\cvir x)/g(\cvir)$, with
\begin{equation}\label{Eq:gc}
g(x)=\begin{cases}
               \ln(\frac{x}{2})+\frac{1}{\sqrt{1-x^2}}\operatorname{arccosh}(\frac{1}{x}),\quad\quad\quad x<1\\
               1+\ln(\frac{1}{2}),\quad\quad\quad\quad\quad\quad\quad\quad\quad\quad x=1\\
               \ln(\frac{x}{2})+\frac{1}{\sqrt{x^2-1}}\operatorname{arccos}(\frac{1}{x}),\quad\quad\quad x>1\\
                \end{cases}
\end{equation}
(e.g., Golse \& Kneib 2002).

The solid black line in the lower left-hand panel of
Fig.\ref{Fig:Fsub} plots the median 3D subhalo mass fraction profile
obtained from 2000 model realizations of haloes with
$M_0=10^{13.1}\Msunh$ at redshift $z_0=0.2$. The dark and light grey
bands indicate the 68\% and 95\% intervals, and reflect the expected
halo-to-halo variance. The other (colored) curves, correspond to
simulations results, as indicated. Since these simulation results
correspond to different host halo masses (Milky Way-size for ELVIS,
Aquarius, and COCO, and cluster-size for Bolshoi and Phoenix), and
since the subhalo mass fraction scales with halo mass (cf.,
Fig.~\ref{Fig:MassFrac}), we re-scale the simulation results to
$M_0=10^{13.1}\Msunh$ using the mass fraction-halo mass relation,
$\fsub\propto M_0^{0.12}$ \footnote{Where necessary we convert halo
  masses and radii to our definition of virial mass and virial radius
  using the concentration-mass relation of Macc\`io \etal
  (2008).}. For ELVIS and Bolshoi we use the halo catalogs at $z=0$
and $z=0.2$, respectively, while for Aquarius, Phoenix and COCO we use
the published results (corresponding to $z=0$) from Gao \etal (2012)
and Bose \etal (2016). Using the Bolshoi simulation results we have
verified that there is no significant evolution in the radial bias
function between $z=0$ and $z=0.2$, justifying the usage of $z=0$
results. The resulting radial subhalo mass fraction profiles are
overall in good mutual agreement. The Aquarius and Phoenix results are
somewhat deviant, in that they give somewhat lower subhalo mass
fractions at small radii. We suspect that these (small) differences
arise mainly as a consequence of using different halo finders.

Finally, the lower right-hand panel of Fig.\ref{Fig:Fsub} plots the
median 2D profile for the same models, and compares them with the
lensing result of Vegetti \etal (2014), represented by the data point
with errorbars.  The $X$-coordinate of the data point is obtained by
dividing $\langle R_\rmE \rangle$ by the median virial radius of the
lenses, $\langle \rvir\rangle \simeq427_{-87}^{+112}\kpch$, i.e., the
virial radius of a halo with $M_0=10^{13.1\pm0.3}\Msunh$, where we
have assumed an uncertainty of $0.3$dex when computing the halo mass
using the Moster \etal (2013) stellar-to-halo-mass relation. As for
the Okabe \etal (2014) results for Coma, the lensing measurement is
much higher than the median of the model predictions or the simulation
results. It does, however, fall with the 95th percentile, as indicated
by the lighter grey band.  Therefore we conclude that the observed
subhalo mass fraction, inferred from the few gravitational lensing
systems investigated so far, is noticeably higher than the median
$\Lambda$CDM prediction. However, taking the large halo-to-halo
variance into account, the level of disagreement does not represent a
significant challenge for the $\Lambda$CDM paradigm.
 
We caution that our model prediction is based on dark matter only
simulations. Baryonic processes can have important effects on the
masses and spatial distribution of subhaloes, especially inside $\sim
1\%$ of the virial radii where the stellar mass of the central galaxy
is likely to contribute significantly to the total enclosed mass.  The
baryonic effects on subhalo statistics is an active area of research,
and there is no concordance yet regarding the extent to which baryonic
physics can suppress or boost the amount of substructure (e.g.,
Despali \& Vegetti 2016; Fiacconi \etal 2016).  We will revisit this
topic in a future study using a suite of high-resolution
hydrodynamical zoom-in simulations.


\section{Halo Occupation Statistics} \label{Sec:sHOD}

Halo occupation distributions (e.g., Berlind \& Weinberg 2002; Zheng
\etal 2005; Giocoli \etal 2010b) and the closely related conditional
luminosity function (e.g., Yang \etal 2003; van den Bosch \etal 2013)
are popular tools for modeling galaxy clustering.  In such models, the
primary quantity of interest is the halo occupation distribution
(HOD), $P(N_{\rm gal}|M_0)$, which expresses the probability for a
halo of mass $M_0$ to host $N_{\rm gal}$ galaxies.  HODs can be used,
in combination with a halo mass function and a halo bias function, to
predict the galaxy-galaxy two-point correlation function $\xi_{\rm
  gg}(r)$, or the associated power spectrum, $\Pgg(k)$ (e.g.,
Cooray \& Sheth 2002; van den Bosch \etal 2013).

The HOD of galaxies is usually decomposed into a central and a
satellite term: $P(N_{\rm gal}|M_0) = P(N_{\rm cen}|M_0) + P(N_{\rm
  sat}|M_0)$. Since satellite galaxies are believed to reside in dark
matter subhaloes, the satellite term is directly related to the
occupation distribution of subhaloes, $P(N_{\rm sub}|M_0)$ (Kravtsov
\etal 2004). In this section we use our semi-analytical model and the
simulations listed in Table~1 to study $P(N_{\rm sub}|M_0)$ in detail,
focusing in particular on its first and second moments. In what
follows we will drop the subscripts `sat' and `sub', and simply use
$N$ to denote the number of satellite galaxies or subhaloes (above a
given mass limit).

The first moment of the HOD is given by $\langle N|M_0 \rangle =
\sum_N N \, P(N|M_0)$, and is needed to compute the 2-halo and 1-halo
central-satellite components of the power spectrum. When doing so, one
implicitly makes the assumption that halo occupation statistics are
completely determined by the mass of the host halo. However, this
assumption is incorrect. After all, as we have shown in
\S\ref{Sec:MassFraczformDependence}, for fixed $M_0$ the subhalo mass
fraction, and hence $\langle N_{\rm sub}|M_0 \rangle$, is strongly
dependent on halo formation time. And since halo bias is also strongly
dependent on halo formation time, an effect known as halo assembly
bias (e.g., Gao \etal 2005, 2007: Wechsler \etal 2006), the clustering
predicted with standard HOD models is systematically, and
significantly, biased (e.g., Zentner \etal 2014).

\begin{figure}
\centerline{\psfig{figure=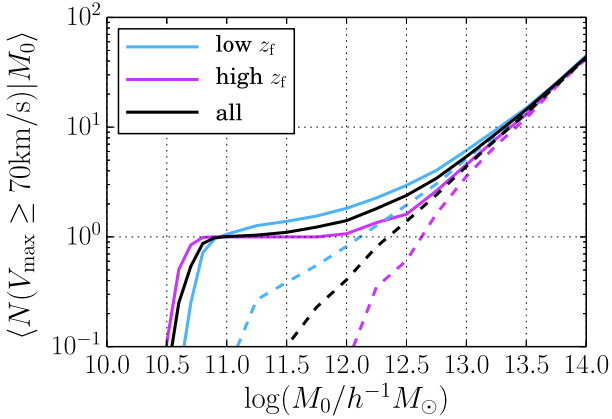,width=0.5\hdsize}}
\caption{Illustration of assembly bias in the occupation number of
  subhaloes. The solid, black lines shows the first moment of the halo
  occupation distribution of haloes with $\Vmax>70\kms$, computed
  using $20,000$ model realizations of haloes with mass uniformly
  distributed in $[10^{10},10^{14}]\Msunh$.  The solid cyan and
  magenta curves show the first moments of the quartiles with the 25\%
  youngest and oldest host haloes, respectively. Dashed lines shows
  the contributions due to subhaloes (i.e., satellite galaxies) only.}
\label{Fig:zformDependentHOD}
\end{figure}

To illustrate how halo formation time impacts the subhalo occupation
distribution (hereafter sHOD), we use our semi-analytical model to
construct $20,000$ model realizations of host haloes with mass
uniformly distributed over $[10^{10},10^{14}]\Msunh$. The solid, black
line in Fig.~\ref{Fig:zformDependentHOD} shows the average number of
haloes with $\Vmax > 70 \kms$, where we also count the host halo. The
dashed, black line shows the corresponding number of subhaloes (i.e.,
the first moment of the sHOD).  At each narrow bin in host halo mass,
we rank-order the host haloes according to their formation redshift,
$\zf$.  The purple and cyan lines show the occupation statistics for
the upper (high $\zf$) and lower (low $\zf$) quartiles,
respectively. Clearly, and in line with the results from
\S\ref{Sec:MassFraczformDependence}, earlier forming haloes host fewer
subhaloes (see also Mao \etal 2015, who obtained very similar results
using numerical simulations). Note also that the full HOD (including
the host haloes) for early-forming haloes extends to smaller masses
than that of its late-forming counterpart. This is simply a
consequence of the fact that earlier forming haloes have higher
concentrations, and therefore larger $\Vmax$.

The second moment of the HOD is given by $\langle N^2 |M_0 \rangle =
\sum_N N^2 \, P(N|M_0)$, and is required to compute the
satellite-satellite component of the 1-halo term. The expectation
value for the number of satellite-satellite pairs in a halo of mass
$M_0$ is
\begin{equation}
  {1 \over 2} \, \langle N(N-1) | M_0 \rangle = {\langle N^2 |M_0 \rangle -
  \langle N|M_0 \rangle \over 2}\,,
\end{equation}
which allows one to compute the 1-halo satellite-satellite term of the
galaxy power spectrum as
\begin{equation} \label{Eq:definePss1h}
P_{\rm ss}^{\rm 1h}(k) = \int_{0}^{\infty} \frac{\langle
  N(N-1)|M_0\rangle}{\bar{n}^2} \, \tilde{u}(k|M_0)^2 \, n(M_0) \, \rmd M_0\,.
\end{equation}
Here $n(M_0)$ is the halo mass function, $\bar{n}=\int \langle
N|M_0\rangle n(M_0) \rmd M_0$ is the average comoving number density,
and $\tilde{u}(k|M_0)$ is the Fourier transform of the radial number
density distribution of satellite galaxies (subhaloes).

It is common practice to assume that the number of subhaloes, and thus
satellite galaxies, follows a Poisson distribution, $P(N|M_0) =
\lambda^{N} {\rm e}^{-\lambda}/N!$, with $\lambda=\langle N|M_0
\rangle$. In that case, $\langle N(N-1)|M_0 \rangle = \langle N|M_0
\rangle^2$, and therefore the first moment of the HOD is sufficient to
compute $P_{\rm ss}(k)$. More generally, we may write
\begin{equation} \label{Eq:definealpha}
\langle N(N-1)|M_0 \rangle = \alpha^2(M_0) \, \langle N|M_0 \rangle^2
\end{equation}
where $\alpha(M_0)$ is a new function that relates the first two
moments of $P(N|M_0)$. Substitution in Eq.(\ref{Eq:definePss1h}) then
yields
\begin{equation}
P_{\rm ss}^{\rm 1h}(k) = \int_{0}^{\infty} \left[ \alpha(M_0) \,
  \frac{\langle N|M_0\rangle}{\bar{n}} \tilde{u}(k|M_0)\right]^2
  \, n(M_0) \, \rmd M_0.
\end{equation}
If subhaloes (satellites) obey Poisson statistics, then $\alpha(M_0) =
1$. Similarly, if $\alpha > 1$ ($< 1$) we say that $P(N|M_0)$ is
super- (sub-)Poissonian. As mentioned above, almost all HOD models
published in the literature assume that satellite galaxies obey
Poisson statistics. Notable exceptions are Cacciato \etal (2013), who
instead assumed that $\alpha$ is independent of mass, so that it can
be taken outside of the integral, and Porciani, Magliocchetti \&
Norberg (2004), who discussed the impact of a mass dependent $\alpha$
on the clustering of quasars.
  
But how accurate is the assumption of a Poissonian $P(N|M_0)$?
Semi-analytic models and hydrodynamic simulations predict a $P(N_{\rm
  gal}|M_0)$ that is significantly sub-Poissonian whenever $\langle
N_{\rm gal}|M_0 \rangle$ is small (e.g., Benson \etal 2000; Seljak
2000; Berlind \etal 2003). In fact, it was realized early on that in
order for HOD models to produce galaxy-galaxy correlation functions
with a power-law-like form as observed, $P(N_{\rm gal}|M_0)$ indeed
needs to be sub-Poissonian (e.g., Benson \etal 2000; Seljak 2000;
Scoccimarro \etal 2001; Berlind \& Weinberg 2002).  However, this does
not necessarily imply that $P(N_{\rm sat}|M_0)$ has to be
sub-Poissonian as well.  In fact, if $P(N_{\rm sat}|M_0)$ is
Poissonian, then as long as $\langle N_{\rm sat} |M_0 \rangle =
\langle N_{\rm gal} | M_0 \rangle - 1$ (which holds as long as $M_0$
is large enough so that the halo always hosts a central galaxy), one
has that
\begin{equation}
  {\langle N_{\rm gal}(N_{\rm gal}-1) | M_0 \rangle \over
    \langle N_{\rm gal} |M_0 \rangle^2} = 1 -
  {1 \over \langle N_{\rm gal}|M_0 \rangle^2}
\end{equation}
(Kravtsov \etal 2004). Hence, $P(N_{\rm gal}|M_0)$ will be
sub-Poissonian unless $\langle N_{\rm gal}|M_0 \rangle$ is large.
Using numerical simulations, Kravtsov \etal (2004) argued that the
occupation distribution of {\it subhaloes}, $P(N_{\rm sub}|M_0)$, is
well described by a Poisson distribution (in that $\alpha \sim 1$),
although they did find evidence for sub-Poissonian behavior at small
$\langle N_{\rm sub}|M_0 \rangle$. More recently, Boylan-Kolchin \etal
(2010, hereafter BK10) pointed out that the sHOD is actually strongly
super-Poissonian when $\langle N|M_0 \rangle$ is large, eventhough
$\alpha \sim 1$ (see also Busha \etal 2011; Wu \etal 2013). This can
come about because one can rewrite Eq.~(\ref{Eq:definealpha}) as
\begin{equation}\label{Eq:alpha}
  \alpha^2(M_0) = 1 + {1 \over \langle N|M_0 \rangle}
  \left( {\sigma^2(M_0) \over \sigma^2_\rmP(M_0)}  - 1\right)\,
\end{equation}
Here $\sigma^2(M_0)$ is the variance of $P(N|M_0)$ and
$\sigma^2_\rmP(M_0)$ is the variance of a Poissonian $P(N|M_0)$ with
the same mean.  Hence, when $\langle N|M_0 \rangle$ is large, even a
strongly non-Poissonian distribution will have a value for $\alpha$
close to unity. Mao \etal (2015) argue that the super-Poissonian
character of the sHOD arises from variance in the large-scale
environments of the host haloes, and that the sHOD is actually
Poissonian for host haloes in a fixed large-scale environment.

In the following subsections, we characterize in detail how the sHOD
deviates from a Poisson distribution. We show that it is both
sub-Poissonian at small $\langle N|M_0\rangle$ and super-Poissonian at
large $\langle N|M_0\rangle$, and present an accurate fitting function
for $\alpha(\Nave)$ (\S\ref{Sec:NonPoisson}). Next we discuss the
implications of the non-Poissonity for galaxy clustering
(\S\ref{Sec:1haloTerm}), and how the detailed shape of the sHOD
depends on how subhaloes are selected (\S\ref{Sec:4types}), Finally,
in \S\ref{Sec:z50Dep} we show that, contrary to the claim by Mao \etal
(2015), non-Poissonity cannot be eliminated by selecting haloes of
fixed formation time. However, we first test to what accuracy our
semi-analytical model can reproduce the detailed shape of the sHOD as
inferred from numerical simulation.


\subsection{Model versus Simulations} \label{Sec:Comparison}

The semi-analytical model has proven to be accurate in predicting the
average subhalo mass and velocity functions (Paper I \& II), as well
as the halo-to-halo variance of the $\vmax$ distributions of massive
subhaloes (Jiang \& van den Bosch \etal 2015).  Here we further
examine the accuracy of the model, focusing on the halo-to-halo
variance in subhalo number, $\langle N(\ge\psi)\rangle$, over a wide
range of subhalo mass, $\psi=m/M_0$.
\begin{figure*}
\centerline{\psfig{figure=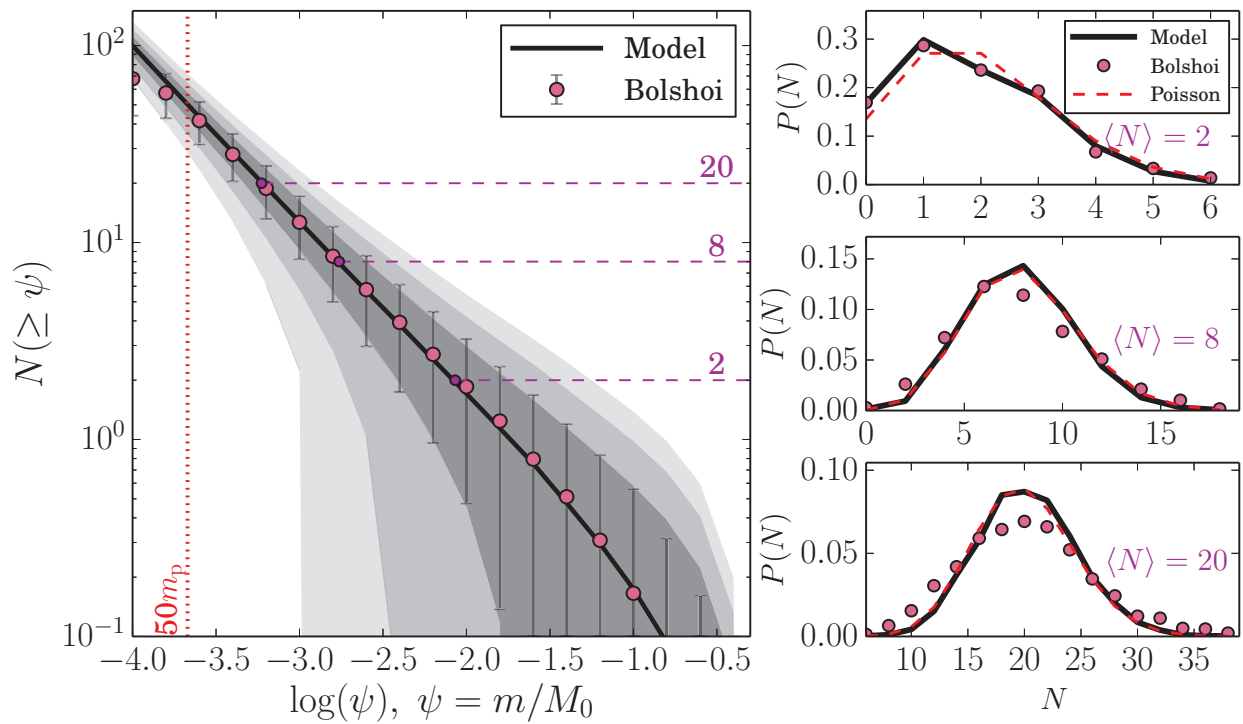,width=0.95\hdsize}}
\caption{Subhalo Occupation Statistics.  {\it Left-hand panel}: the
  average, cumulative subhalo mass function, $\langle
  N(\ge\psi)\rangle$, for 10,000 model realizations of haloes with
  $M_0=10^{13.75}\pm0.25\Msunh$ (solid, black line), compared with
  that of the 1231 Bolshoi haloes in the same mass range (symbols).
  The progressively lighter grey bands indicate the $\pm1$, 2, and
  3$\sigma$ scatter around the average, while the errorbars indicate
  the $\pm1\sigma$ scatter in Bolshoi, where $\sigma=\sqrt{{\rm
      Var}(N)}$.  The vertical dotted line indicates the Bolshoi
  resolution limit of $50m_\rmp$.  {\it Right-hand panels}: subhalo
  occupation distributions, $P(N(\ge\psi)|M_0)$, for different mass
  thresholds, $\psi$, for which $\Nave=2$, 8, and 20 (as
  indicated). The model prediction (solid line) starts to disagree
  with the Bolshoi results (symbols) as $\Nave$ increases. The red,
  dashed curves correspond to Poisson distributions with the same
  $\Nave$, and are shown for comparison.  See text for a detailed
  discussion.}
\label{Fig:Comparison}
\end{figure*}

We consider 10,000 model realizations of haloes with
$M_0=10^{13.75\pm0.25}\Msunh$, and the 1231 Bolshoi haloes in the same
mass range.  The left-hand panel of Fig.\ref{Fig:Comparison} compares
$\langle N(\ge\psi)\rangle$ of the model realizations, to that of the
Bolshoi haloes.  The model is generally in good agreement with the
simulation in terms of the average, as well as the halo-to-halo
variance at the massive end, but it underestimates the scatter at the
low-mass end.  This is better revealed in the right-hand panels of
Fig.\ref{Fig:Comparison}, which compare the model and Bolshoi sHODs,
for three different mass thresholds at which $\langle N(\ge\psi)
\rangle=$ 2, 8, and 20 respectively. For $\Nave = 2$ the Bolshoi
simulation results reveal a clear departure from a Poisson
distribution, which is accurately reproduced by the model. Hence, the
semi-analytical model seems to reproduce the sub-Poissonian behavior
of the sHOD for small $\Nave$. However, when $\Nave$ becomes large,
the sHOD from the Bolshoi simulation becomes super-Poissonian,
something that is {\it not} captured by the semi-analytical model,
which instead predicts sHODs that are perfectly Poissonian.

This failure of the semi-analytical model to reproduce the strongly
super-Poissonian nature of the sHOD, is most likely due to the fact
that the halo merger trees in the model are generated using the
Parkinson \etal (2008) algorithm, which is based on the extended
Press-Schechter (EPS) theory.  EPS describes the assembly of dark
matter haloes using Markovian excursion sets, which means that halo
mass increments are assumed to be uncorrelated. As shown by Neistein
\& Dekel (2008), this is not representative of how haloes assemble in
numerical simulations, and explains why the standard (i.e., Markovian)
EPS formalism fails to capture the non-Poissonity of subhalo
statistics {\it accretion}.  We conclude that despite the model's
successes, it is not suitable for predicting the detailed shape of the
sHOD (at least not for $\Nave \gta 2$). Hence, in what follows we
primarily use the Bolshoi and MultiDark simulations to study the
non-Poissonity of $P(N|M_0)$.
\begin{figure}
\centerline{\psfig{figure=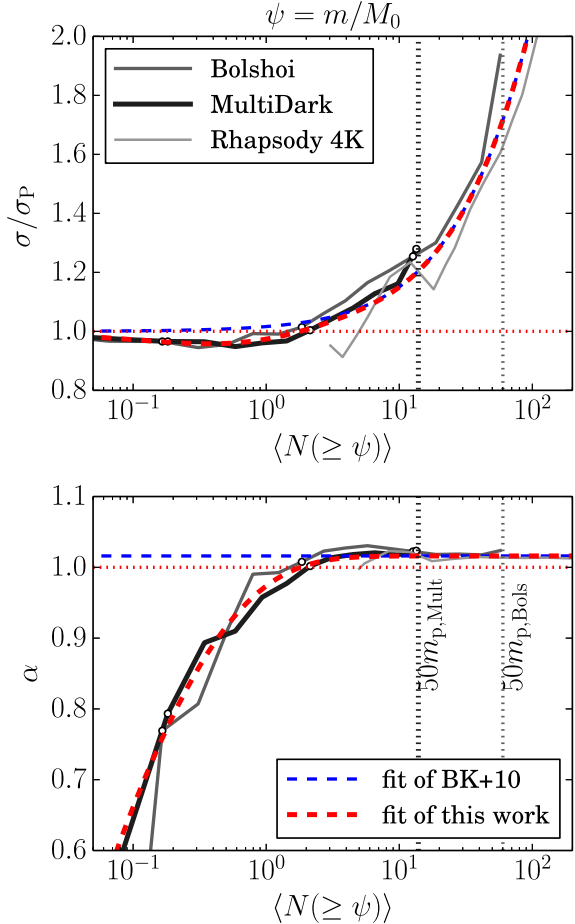,width=0.45\hdsize}}
\caption{ {\it Upper panel}: halo-to-halo variance of subhalo number,
  $\sig=\sqrt{{\rm Var}(N)}$, in units of the corresponding Poisson
  scatter, $\sigP=\sqrt{\Nave}$, as a function of the average
  cumulative number of subhaloes, $\langle N(\ge\psi)\rangle$.  Shown
  in solid lines are the Bolshoi result for the 1231 haloes with
  $M_0=10^{13.75\pm0.25}\Msunh$, the 2393 MultiDark result for haloes
  with $M_0=10^{14.75\pm0.25}\Msunh$, and the 96 Rhapsody haloes with
  $M_0=10^{14.8\pm0.05}\Msunh$, as indicated.  The cyan, dashed line
  indicates the model of BK10 based on the MS-II simulation
  (Eq.~[\ref{Eq:BK10}]), while the red dashed line is the new fitting
  function proposed here (Eq.~[\ref{Eq:sigsigP}]).  Vertical, dotted
  lines indicate the 50-particle resolution limits of the Bolshoi and
  MultiDark simulations, as indicated. {\it Lower panel}: Same as the
  upper panel, but for $\alpha \equiv \langle N(N-1)\rangle^{1/2} /
  \langle N \rangle$, which is the parameter relevant for HOD modeling
  of galaxy clustering.}
\label{Fig:NonPoissonFit}
\end{figure}
%


\subsection{Non-Poissonity} \label{Sec:NonPoisson}

The upper panel of Fig.\ref{Fig:NonPoissonFit} plots the standard
deviation of the subhalo number counts, $\sig=\sqrt{{\rm Var}(N)}$, in
units of that of a Poisson distribution, $\sigP=\sqrt{\Nave}$, as a
function of the average subhalo number, $\Nave$.  Here we use the 1231
Bolshoi haloes with $M_0=10^{13.75\pm0.25}\Msunh$ and the 2393
MultiDark haloes with $M_0=10^{14.75\pm0.25}\Msunh$.  The Bolshoi and
MultiDark results are plotted up to $\langle N(\ge\psi_{\rm res})
\rangle$, where $\psi_{\rm res}=\mres/M_0\simeq10^{-3.7}$ and
$10^{-2.9}$ respectively, with $\mres$ the mass of 50 particles.  The
halo mass scales are chosen as a compromise between sample size and
dynamical range: more massive haloes are better resolved than
lower-mass haloes, thereby allowing one to probe the behavior to
larger $\Nave$. On the other hand, one needs of the order of $1000$
haloes to reliably probe the statistics at small $\Nave$, and the
number density of massive haloes is limited. We have verified, though,
using haloes of different mass scales in Bolshoi and MultiDark, that
the results exhibit little dependence on halo mass.  Also plotted, for
comparison, are the published results for the 96 haloes in the
`Rhapsody-4K' simulation with $M_0=10^{14.8\pm0.05}\Msunh$ (Wu \etal
2013).  Taken together, these simulation results reveal that the
scatter in $P(N_{\rm sub}|M_0)$ is super-Poissonian ($\sig/\sigP>1$)
when $\Nave$ is large, and sub-Poissonian ($\sig/\sigP<1$) when
$\Nave$ is small, with the transition occurring at $\Nave\sim2$.

The sub-Poissonian scatter at small $\Nave$ is better revealed in the
lower panel of Fig.\ref{Fig:NonPoissonFit}, which plots the
$\alpha$-statistic (Eq.~[\ref{Eq:alpha}]) as a function of $\Nave$, At
large $\Nave$, $\alpha$ in different simulations converge to a
constant that is slightly larger than unity, while at small $\Nave$,
$\alpha$ drops below unity.

The super-Poissonian behavior at large-$\Nave$ is well known.  Based
on the results from the Millennium-II simulation (hereafter MS-II),
BK10 describe the halo-to-halo variance as the sum of two terms,
$\sig^2 = \sigP^2+\sigI^2$.  The first term is the Poissonian scatter,
$\sigP^2 = \Nave$; while the second term reflects {\it intrinsic}
scatter. Fitting to the MS-II results, they find that, to good
approximation, $\sigI^2 = (0.18\Nave)^2$.  Therefore, $\sig/\sigP$ can
be expressed as
\begin{equation} \label{Eq:BK10}
\frac{\sig}{\sigP} =  \sqrt{1+\eps^2\Nave},
\end{equation}
with $\epsilon=0.18$.  As shown in Fig.\ref{Fig:NonPoissonFit} (blue
dashed curve), this simple, empirical model also nicely fits the
Bolshoi and MultiDark results at $\Nave \gg 2$.  However, it does not
capture the sub-Poissonian scatter at $\Nave\la 2$: according to
Eqs.(\ref{Eq:BK10}) and~(\ref{Eq:alpha}), the BK10 model yields
$\alpha=\sqrt{1+\eps^2}=1.016$, independent of $\Nave$.  We emphasize
that our results are not inconsistent with the MS-II simulation.
Rather, the latter simply lacks sufficient statistical power to probe
the behavior of $\sig/\sigP$ at small $\Nave$. In fact, Fig.~8 in BK10
already hints at sub-Poissonian behavior at small $\Nave$, at least
for the more massive host haloes, at low significance.

We find that the sub-Poissonian scatter can be taken into account by
multiplying the expression for $\sig/\sigP$ of Eq.(\ref{Eq:BK10}) with
a `correction factor' according to
\begin{equation} \label{Eq:sigsigP}
  \frac{\sig}{\sigP} = \left(1-\eta \, x^2 \, {\rm e}^{-x} \right)
  \sqrt{1+\eps^2\Nave},
\end{equation}
where $x=\sqrt{\Nave/N_0}$, with $\eta=0.09$, $N_0=0.12$, and
$\eps=0.18$. As is evident from Fig.\ref{Fig:NonPoissonFit} (red
dashed curve), this model accurately describes the simulation results
over the entire range of $\Nave$ covered. In particular, for
$\Nave\gg1$, it reduces to Eq.(\ref{Eq:BK10}), such that the
asymptotic value of $\alpha$ is still $\sqrt{1+\eps^2}$, in agreement
with BK10.
\begin{figure*}
\centerline{\psfig{figure=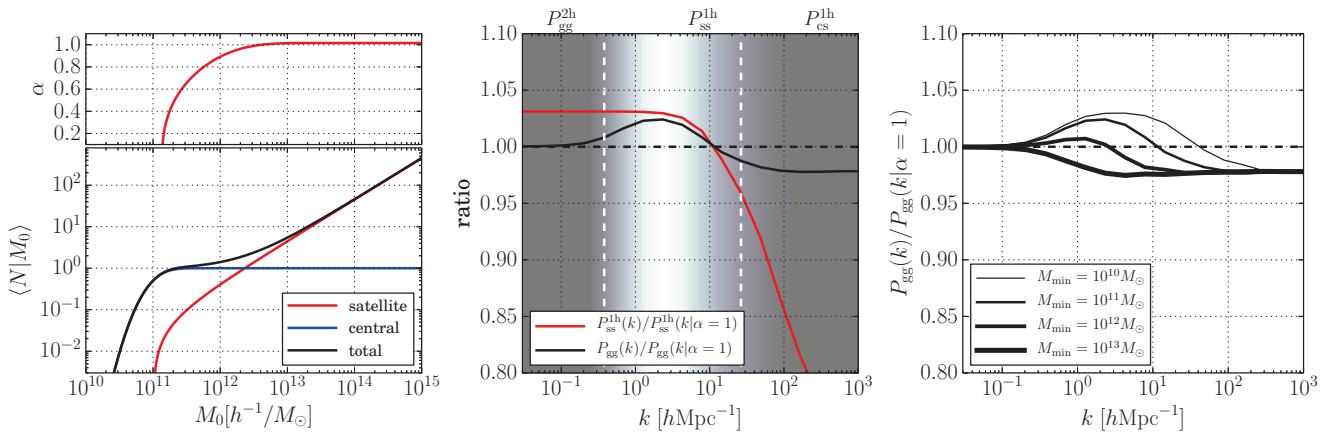,width=\hdsize}}
\caption{ Illustration of the impact of non-Poissonian subhalo
  statistics on the galaxy power spectrum.  {\it Left-hand panel}: The
  lower panel shows the first moment of the occupation statistics of
  satellites (red), centrals (blue), and all galaxies (black) for our
  HOD with $\Mmin=10^{11}\Msunh$, which roughly corresponds to
  $\Vmax=70\kms$.  The upper panel plots the corresponding $\alpha$,
  computed using our fitting function as described in the text.  {\it
    Middle panel}: the ratio of the galaxy power spectrum computed
  using the $\alpha(M_0)$ relation given in the upper left-hand panel,
  to that computed using $\alpha = 1$ (as is commonly done). Red and
  black curves correspond to the one-halo satellite-satellite term,
  $\Pss^{1h}(k)$, and the full power spectrum, $P_{\rm gg}(k)$,
  respectively, while the grey scale reflects the fractional
  contribution of $\Pss^{1h}$ to the total power $\Pgg$ (lighter grey
  tones indicating a larger contribution).  The vertical lines bracket
  the regime where $\Pss^{1h}$ contributes more than 50\% of the total
  power.  {\it Right-hand panel}: Same as the solid curve in the
  middle panel, but for different values of $\Mmin$, as
  indicated. Note how ignoring the non-Poissonity introduces
  systematic errors of up to $\sim 3$ percent, and with a complicated
  scale- and $\Mmin$-dependence.}
\label{Fig:HaloModel}
\end{figure*}
%


\subsection{Implications for Satellite Clustering} \label{Sec:1haloTerm}

Since $\alpha$ is a function of $\Nave$, and $\Nave$ depends on halo
mass $M_0$, we have that $\alpha = \alpha(M_0)$. Here we illustrate
the influence of a mass-dependent $\alpha$ on the galaxy power
spectrum. In particular, we show how the non-Poissonity as described
by Eq.~(\ref{Eq:sigsigP}) impacts the galaxy power spectrum $P_{\rm
  gg}(k)$, but comparing with the case were $\alpha=1$, as is commonly
assumed.

Splitting the galaxy population in centrals and satellites, and
ignoring (for simplicity) shot-noise, we can write the galaxy power
spectrum as
\begin{equation} \label{Eq:Pgg}
  \Pgg(k) = 2 \, \Pcs^{\rm 1h}(k) + \Pss^{\rm 1h}(k) + \Pcc^{\rm 2h}(k)
  + 2 \, \Pcs^{\rm 2h}(k) + \Pss^{\rm 2h}(k)\,,
\end{equation}
(e.g., van den Bosch \etal 2013). Here the super-scripts `1h' and `2h'
refer to the one- and two-halo term, respectively, while `c' and `s'
refer to centrals and satellites.

Following van den Bosch \etal (2013), and ignoring halo exclusion and
scale dependence of the halo bias, we have that
\begin{equation} \label{Eq:Pcs1h}
P_{\rm cs}^{\rm 1h}(k) = \int_{0}^{\infty} \calH_\rmc(k,M_0) \,
\calH_\rms(k,M_0) \, n(M_0) \, \rmd M_0.
\end{equation}
\begin{equation} \label{Eq:Pss1h}
P_{\rm ss}^{\rm 1h}(k) = \int_{0}^{\infty} \alpha^2(M_0) \,
\calH^2_\rms(k,M_0) \, n(M_0) \, \rmd M_0.
\end{equation}
and
\begin{eqnarray} \label{Eq:P2h}
P_{\rm xy}^{\rm 2h}(k) & = & P_{\rm lin}(k) \, 
\int_{0}^{\infty} \calH_\rmx(k,M_0) \,  n(M_0) \, b_{\rm h}(M_0) \, \rmd M_0 \nonumber \\
& & 
  \int_{0}^{\infty} \calH_\rmy(k,M_0) \,  n(M_0) \, b_{\rm h}(M_0) \, \rmd M_0 \,.
\end{eqnarray}
Here, $\rmx$ and $\rmy$ are either $\rmc$ (for central) or $\rms$ (for satellite),
$n(M_0)$ is the halo mass function, $b_\rmh(M_0)$ is the linear halo
bias function, and we have used the shorthand notation
\begin{equation}
\calH_\rmc(k,M_0) = {\langle N_{\rm cen}|M_0 \rangle \over \bar{n}} 
\end{equation}
and
\begin{equation}
\calH_\rms(k,M_0) = {\langle N_{\rm sat}|M_0 \rangle \over \bar{n}} \, 
\tilde{u}(k|M_0)
\end{equation}
with $\bar{n}$ the total number density of all galaxies (centrals plus
satellites), and $\tilde{u}(k|M_0)$ the Fourier transform of the
radial number density distribution of satellite galaxies in a halo of
mass $M_0$. Note that centrals are assumed to reside at rest at the
center of their host halo.

In what follows, we adopt the halo mass and halo bias functions of
Tinker \etal (2008, 2010) for haloes that are defined to have an
overdensity with respect to the average matter density of $\Delta_\rmm
= 360$\footnote{Using interpolation of the mass and bias functions
  presented in Tinker \etal for different values of
  $\Delta_\rmm$.}. In addition, we assume that satellites follow a
radial number density distribution given by an NFW profile with the
concentration-mass relation of Neto \etal (2007). The corresponding
expression for $\tilde{u}(k|M_0)$ is given in Scoccimarro \etal
(2001).  Finally, we adopt the following HOD model for the centrals
and satellites:
\begin{equation} \label{Eq:NMc}
\langle N_{\rm cen}|M_0\rangle = {1 \over 2} \left[ 1+{\rm erf}[ (\log M_0-\log
  \Mmin)/0.3 ]\right]
\end{equation}
and
\begin{equation} \label{Eq:NM}
\langle N_{\rm sat}|M_0\rangle =
\begin{cases}
    0.045(\frac{M_0}{\Mmin} - 1), & M_0>\Mmin \\
    0,                            & M_0<\Mmin
\end{cases}
\end{equation}
with $\Mmin = 10^{11}\Msunh$. There is no particular reason for
picking this HOD model other than that it is representative of typical
HOD models used in the literature.  The second moment of the satellite
occupation statistics is characterized by $\alpha(M_0)$, which is
given by the combination of Eqs.~(\ref{Eq:alpha}), (\ref{Eq:sigsigP}),
and~(\ref{Eq:NM}).

The left-hand panels of Fig.\ref{Fig:HaloModel} plot $\langle
N|M_0\rangle$ and $\alpha(M_0)$ corresponding to this HOD model.  The
red, solid curve in the middle panel of Fig.\ref{Fig:HaloModel} plots
the ratio between $P_{\rm ss}^{\rm 1h}(k)$ and $P_{\rm ss}^{\rm
  1h}(k|\alpha=1)$. Note how the $M_0$-dependence of $\alpha$ reduces
the clustering power at small scales (large-$k$) while enhancing it
(by about 3 percent) on large scales (small-$k$).  The grey-scale
reflects the fractional contribution of this satellite-satellite
1-halo term to the total power spectrum, $P_{\rm gg}(k)$, with a
darker grey-tone reflecting a small contribution: on large scales the
2-halo term dominates, while the central-satellite 1-halo term
dominates on small scales. The black, solid curve in the middle panel
of Fig.\ref{Fig:HaloModel} plots the ratio between $P_{\rm gg}(k)$ and
$P_{\rm gg}(k|\alpha=1)$, which is basically unity on large scales
(small-$k$). However, for $k \gta 1 h {\rm Mpc}^{-1}$ ignoring the
non-Poissonity of the sHOD can result in systematic errors in the
galaxy power-spectrum of up to $\sim 3$ percent. In particular, it
will underpredict the power by $\sim 3$ percent on intermediate scales
($k \sim 3 h {\rm Mpc}^{-1}$), and overpredict the power by $\sim 2$
percent on very small scales ($k > 100 h {\rm Mpc}^{-1}$). Finally,
the right-hand panel of Fig.\ref{Fig:HaloModel} plots $P_{\rm gg}(k) /
P_{\rm gg}(k|\alpha=1)$ for four different values of $M_{\rm min}$, as
indicated. Roughly speaking, lower values of $M_{\rm min}$ corresponds
to HODs of fainter galaxies. As is evident, the non-Poissonity of the
sHOD manifests itself differently for different luminosity-threshold
samples of galaxies; typically, the power spectra of fainter galaxies
are more susceptible to the detailed shape of the sHOD (i.e., show a
larger deviation from the case with $\alpha = 1$).  Hence, simply
making the naive assumption that the occupation statistics of
satellite galaxies are Poissonian results in systematic errors in the
galaxy-power spectrum (and thus also in the corresponding two-point
correlation function) at the level of a few percent, and with a
complicated scale- and luminosity-dependence. Although small, such
errors are funest for any attempt to use galaxy clustering to do
precision (i.e. percent-level) cosmology.
\begin{figure}
\centerline{\psfig{figure=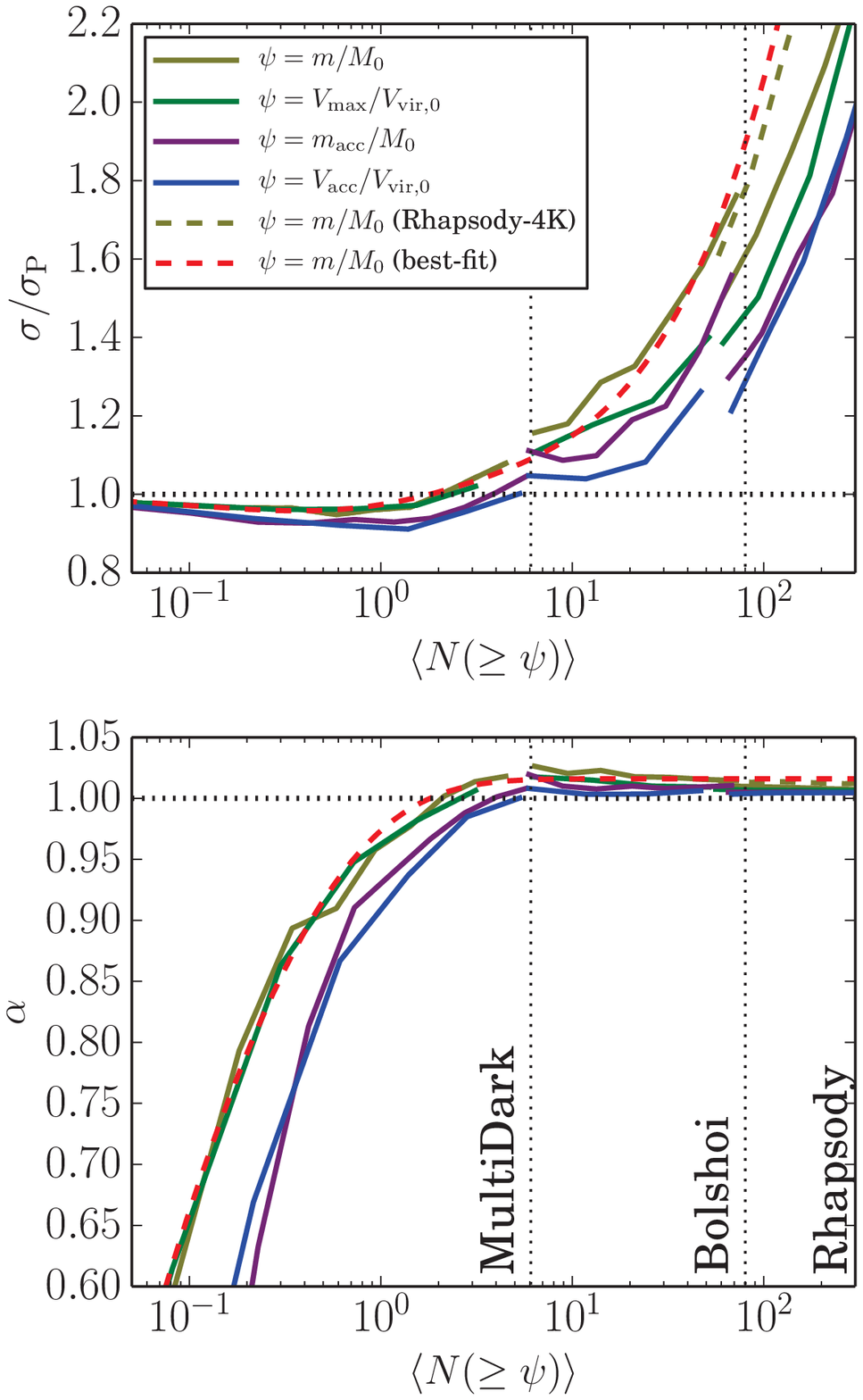,width=0.45\hdsize}}
\caption{ Dependence of non-Poissonity on subhalo selection.  The
  upper and lower panels plot, respectively, $\sig/\sigP$ and $\alpha$
  as functions of $\langle N(\ge\psi)\rangle$, for $\psi=m/M_0$,
  $\vmax/\Vvir$, $\macc/M_0$, and $\vacc/\Vvir$.  From small-$\Nave$
  to large-$\Nave$, we use the results for MultiDark haloes with
  $M_0=10^{14.75\pm0.25}\Msunh$, Bolshoi haloes with
  $M_0=10^{14.25\pm0.25}\Msunh$, and Rhapsody haloes with
  $M_0=10^{14.8\pm0.05}\Msunh$, respectively.  The two vertical dotted
  lines indicate the mass scales where subhaloes have 100-particles in
  the MultiDark and Bolshoi simulation, as indicated.  The red, dashed
  lines correspond to the same fitting function as shown in
  Fig.\ref{Fig:NonPoissonFit}, which best describes the $m$-selected
  samples. Using a selection based on $\macc$ or $\vacc$ yields weaker
  super-Poissonity at high-$\Nave$ and stronger sub-Poissonity at
  low-$\Nave$, compared to the selection based on $m$ or $\vmax$.  The
  dashed, ochre line is result based on $m$-selection for the
  Rhapsody-4K simulation.  It reveals stronger super-Poissonity than
  for the higher-resolution, fiducial Rhapsody-8K simulation,
  suggesting that the simulation results have not yet fully converged.}
\label{Fig:NonPoisson4types}
\end{figure}
%


\subsection{Dependence on Subhalo Selection} \label{Sec:4types}

Wu \etal (2013) found that selecting subhaloes by their {\it
  present-day} mass or $\vmax$ results in stronger super-Poissonity
than selecting by their mass or $\vmax$ {\it at accretion}.  Their
result is based on the relatively small sample of 96 Rhapsody haloes
and limited to the regime where $\Nave\ga100$. We now revisit this
issue using much larger samples from the Bolshoi and MultiDark
simulations, with the goal to characterize in detail how
$\alpha(\Nave)$ depends on how subhaloes are selected.

Fig.\ref{Fig:NonPoisson4types} plots $\sig/\sigP$ and $\alpha$ as
functions of $\langle N(\ge\psi) \rangle$ for different choices of
$\psi$.  The MultiDark, Bolshoi, and Rhapsody results are plotted for
small-, intermediate-, and large-$\Nave$, respectively.  The red,
dashed lines represent the same fitting functions as shown in
Fig.\ref{Fig:NonPoissonFit}, which accurately describe the results for
$\psi=m/M_0$ and serve as reference lines to facilitate the
comparison. A few trends are evident.  First, confirming the Wu \etal
(2013) result, we find weaker super-Poissonity for $\macc$- and
$\vacc$-selected subhaloes. However, at small $\Nave$, $\macc$- or
$\vacc$-selection results in a stronger sub-Poissonity compared to a
selection based on $m$ or $\vmax$. The $\macc$- and $\vacc$-selected
samples behave extremely similar; both have a $\sig/\sigP$-$\Nave$
dependence that is well approximated by Eq.(\ref{Eq:sigsigP}) with
$\eta=0.16$, $N_0=0.2$, and $\eps=0.1$. Second, whereas the results
for MultiDark and Bolshoi nicely agree around their transition at
$\Nave = 5$, this is not the case for Bolshoi and Rhapsody, which
reveal a pronounced `jump' in $\sig/\sigP$ around their transition at
$\Nave = 80$, especially for $m$ and $\macc$.  Interestingly, though,
the result from the lower-resolution Rhapsody-4K, shown as the dashed,
green-yellow line for $\psi=m/M_0$ only, matches the Bolshoi results
extremely well.  This suggests that the super-Poissonity at
large-$\Nave$ may be sensitive to resolution effects in the numerical
simulation. Hence, the we caution that the model for $\alpha(\Nave)$
presented in \S\ref{Sec:NonPoisson} is only as good as the simulations
used, and may be impacted by numerical resolution effects.
\begin{figure*}
\centerline{\psfig{figure=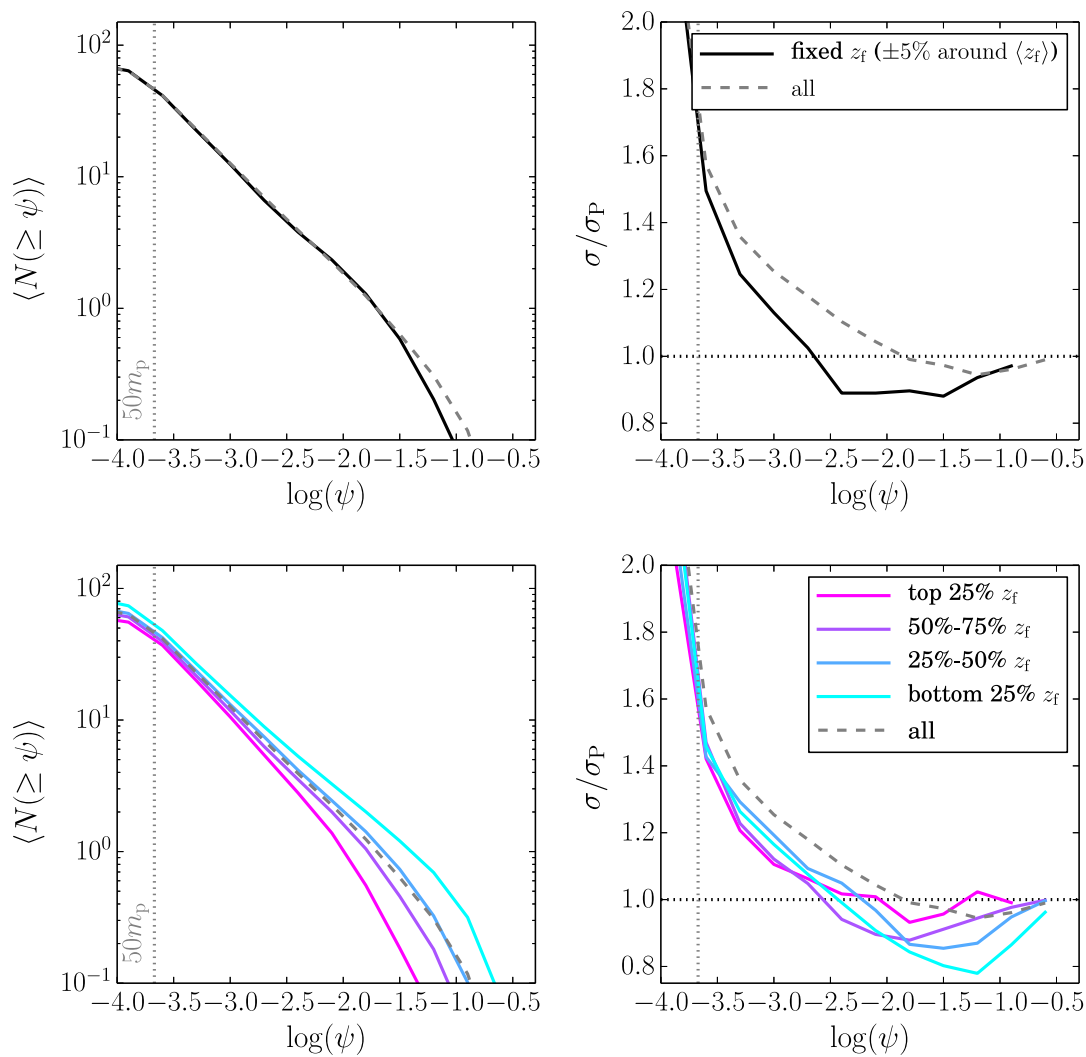,width=0.85\hdsize}}
\caption{ {\it Upper left-hand panel}: The grey, dashed curve shows
  the average, cumulative mass function, $\langle N(\ge\psi) \rangle$,
  for all haloes in the Bolshoi simulation with
  $M_0=10^{13.75\pm0.25}\Msunh$. The solid, black line corresponds to
  the subset of host haloes that have formation redshifts in a narrow
  ($\pm 5$\% percentile) range centered around the median $\zf$.  {\it
    Upper right-hand panel}: the corresponding ratios
  $\sig/\sigP$. Note how restricting $\zf$ to a narrow range the
  super-Poissonity becomes weaker while the sub-Poissonity becomes
  stronger. {\it Lower panels:} same as upper panels, but this time
  showing results for the four different quartiles in formation
  redshift, as indicated. Note how older host haloes have subhalo
  occupation statistics that are closer to Poissonian.}
\label{Fig:NonPoissonESMFinFormationTimeBins}
\end{figure*}
%


\subsection{Environmental Effects} \label{Sec:z50Dep}

Mao \etal (2015) suggest that the non-Poissonity of the sHOD
originates from the large-scale environment of the host haloes.  In
particular, they argue that host haloes of the same mass and with the
same large scale environment have a Poissonian sHOD, and that the
super-Poissonity is a consequence of `convolving' this Poissonian sHOD
with a distribution of environments, each having a slightly different,
albeit Poissonian, sHOD.  Based on this notion, they devise a simple
model for the subhalo $\vmax$ function, in which the {\it average}
subhalo abundance, $\langle N(\ge\vmax) \rangle$, is a function of the
ratio $\Vmax/\Vvir$ of the host halo. In addition, for a given
$\Vmax/\Vvir$ it is assumed that $N(\ge\vmax)$ follows Poisson
statistics.  Hence, Mao \etal (2015) treat $\Vmax/\Vvir$ as their
environment proxy, which is motivated by the fact that $\Vmax/\Vvir$
depends on halo concentration, which depends on the halo formation
time, $\zf$, which in turn depends on the halo's large scale
environment (e.g., Wechsler \etal 2002; Hearin, Behroozi \& van den
Bosch 2016).

In order to test this claim by Mao \etal (2015), we investigate
whether the non-Poissonity of the sHOD diminishes if we select host
haloes based on both mass and formation redshift $\zf$; i.e., we treat
halo formation time as a proxy for a host halo's large scale
environment. As a first test, we select a subsample of the 1231
Bolshoi host haloes with $M_0=10^{13.75\pm0.25}\Msunh$ that have a
formation redshift, $\zf$, in the very narrow range, $\zf=0.71 \pm
0.3$. This corresponds to the $\pm 5$ percentile range centered around
the median formation redshift. The upper left-hand panel of
Fig.\ref{Fig:NonPoissonESMFinFormationTimeBins} plots the average,
cumulative subhalo mass functions, $\langle N(\ge\psi) \rangle$, for
all subhaloes in the mass bin, and for those in the $\pm 5$ percentile
range centered on the median $\zf$. As expected, choosing haloes with
$\zf$ around the median $\zf$ has little impact on the average subhalo
mass function. However, as is evident from the upper right-hand panel,
the variance of the `fixed-$\zf$' subsample is fairly different from
that of the full sample. In particular, the super-Poissonity at the
low-mass end is weaker, while the sub-Poissonity at the massive end is
enhanced. Put differently, the transition from super-Poissonian to
sub-Poissonian shifts from $\psi=m/M_0\simeq0.016$ to
$m/M_0\simeq0.002$ (or equivalently from $\Nave\simeq2$ to
$\Nave\simeq5$).  Although not shown here, we have verified that
qualitatively similar behavior is observed if we adopt halo
concentration as the environment proxy and select a subsample of
haloes with fixed $\cvir$.

In order to portray more clearly how $\sig/\sigP$ depends on formation
time, we bin the Bolshoi haloes with $M_0=10^{13.75\pm0.25}\Msunh$
into $\zf$ quartiles and plot their average subhalo mass functions
$\langle N(\ge\psi)\rangle$ and variance ratio $\sig/\sigP$ in the
lower panels of Fig.\ref{Fig:NonPoissonESMFinFormationTimeBins}.  As is
evident from the lower left-hand panel, and as discussed in detail in
\S\ref{Sec:MassFraczformDependence}, older haloes (i.e., those with
higher $\zf$) contain more subhaloes than younger haloes of the same
mass. The difference is most pronounced at the massive end, consistent
with the results of Gao \etal (2011) based on the MS-II simulation.
In terms of the $\sig/\sigP$, there is a clear trend that the sHOD of
younger host haloes deviates more strongly from a Poisson
distribution. Since older haloes acquire their subhaloes earlier and
tend to be more dynamically relaxed, this suggests that the dynamical
evolution of subhaloes (mass stripping and tidal disruption) drives
the sHOD towards a Poisson distribution.

To summarize, we agree with Mao \etal (2015) that the detailed shape
of the sHOD depends on the large scale environment of the host haloes,
at least in as far as this environment is correlated with halo
formation time. The main trend is for the sHOD to become more
Poissonian with increasing halo formation time. However, we see no
indication to support their claim that the sHOD is Poissonian for a
fixed environment. Consequently, their model for $\langle N(\ge\vmax)
\rangle$ is likely to be oversimplified. Mao \etal (2015) based their
conclusion on a suite of 13 zoom-in simulations of a Milky Way-size
halo. Each simulation has the same large scale modes (representative
of the large scale environment), but differ in the initial phases of
the small-scale modes with $k \ga 16.4h/\Mpc$ (comoving). They find
that the subhaloes in this suite obey Poisson statistics (i.e.,
$\sig/\sigP \simeq 1$) when $\langle N(\ge\psi) \rangle$ is large.
However, the Lagrangian volume within which the initial conditions are
randomized is very small compared to that of entire host halo: $k =
16.4h/\Mpc$ corresponds to a scale of $\lambda \simeq 2\pi/k \simeq
0.38\mpch$, while a Milky Way-size halo ($M_0\sim10^{12}\Msunh$)
corresponds to a Lagrangian volume of radius $r_\rmL=[3M_0 /
  (4\pi\rho_\rmm)]^{1/3} \sim 1.44\mpch$, which is more than three
times larger than $\lambda$. In other words, the mass fraction of the
host halo for which Mao \etal randomize the modes is only
$(\lambda/r_\rmL)^3 \simeq 0.019$. Consequently, these 13 zoom-in
simulations have virtually identical mass accretion histories, only
differing in their accretion of low mass subhaloes, whose combined
mass is less than 2 percent of that of the host halo. This is not a
fair assessment of the impact of environment, and explains why Mao
\etal obtained a sHOD that is close to Poissonian for large $\Nave$.
\begin{figure*}
\centerline{\psfig{figure=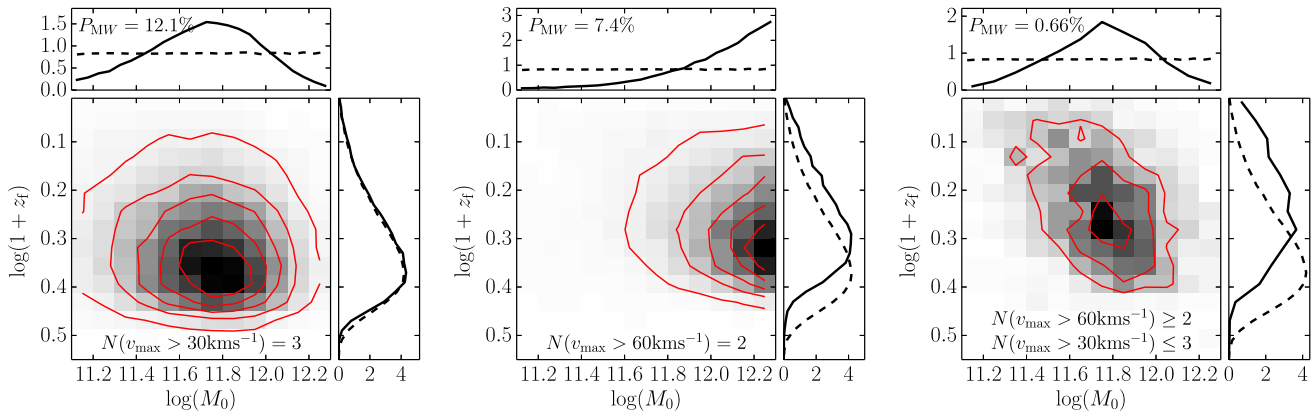,width=\hdsize}}
\caption{The joint distribution of halo mass and formation redshift
  for haloes with exactly 3 massive satellites with $\Vmax>30\kms$
  ({\it left-hand panel}), 2 Magellanic Cloud-analogs with
  $\Vmax>60\kms$ ({\it middle panel}), and with the combined
  requirement that there are no less than 2 Magellanic Cloud-analogs
  and no more than 3 massive satellites ({\it right-hand panel}).
  Results are based 500,000 model realizations of host haloes with
  mass uniformly distributed over the range
  $M_0=10^{11.7\pm0.6}\Msunh$.  The side panels show the marginalized
  distributions of host halo mass (up) and formation redshift (side),
  with solid lines indicating the distributions for haloes that match
  the required occupation condition, and dashed lines representing the
  unconstrained distribution for all 500,000 host haloes.  The $P_{\rm
    MW}$ value printed in the upper left corner of each panel
  indicates the fraction of all host haloes that meet the
  corresponding occupation condition.}
\label{Fig:MW}
\end{figure*}
%


\section{The most massive satellites of Milky Way-size haloes}
\label{Sec:MostMassiveSatellites}

The HOD of massive subhaloes takes central stage in the
too-big-to-fail problem (e.g., Boylan-Kolchin \etal 2011;
Garrison-Kimmel \etal 2014b, Cautun \etal 2014, Jiang \& van den Bosch
2015) and provides one of the most stringent tests of the $\Lambda$CDM
paradigm. In this section, we adopt the cosmological parameters of the
Bolshoi simulation, and use our semi-analytical model combined with
data on the most massive satellite galaxies of the Milky-Way, to
constrain the mass and formation redshift of the Milky-Way's host halo
(see Cautun \etal 2014 for a similar approach based on the MS-II
simulation).

More specifically, we construct 500,000 realizations of host haloes with
mass uniformly distributed in the interval $[10^{11.1},
  10^{12.3}]\Msunh$, and select the realizations that match one or
more of the following three aspects of the satellite occupation
statistics of the Milky Way:
\begin{enumerate}
\item There are three satellites with $\vmax\ga30\kms$.
\item There are two Magellanic Cloud analogues, which have $\vmax\ga60\kms$.
\item There are no less than two Magellanic Cloud analogues and no more than three satellites with $\vmax\ga30\kms$.
\end{enumerate}
The last condition reflects the $\vmax$ `gap' of the Milky Way
satellites, in keeping with Cautun \etal (2014) and Jiang \& van den
Bosch (2015).  The main panels of Fig.\ref{Fig:MW} plot the joint
distributions of host halo mass and formation redshift for haloes that
meet the aforementioned occupation conditions.  The side panels show
the marginalized distributions for haloes that match the corresponding
occupation condition (solid lines), and compare them to the
unconstrained distributions for all the 500,000 haloes (dashed lines).
Any difference between the solid line and the dashed line reflects the
constraint due to the occupation condition.

As shown in the left-hand panel of Fig.~\ref{Fig:MW}, Condition (i)
favors a relatively low Milky Way mass, with the marginalized
distribution peaking at $M_{12} \equiv M_0/10^{12}\Msunh \simeq 0.55$,
and $0.2 < M_{12} < 1.3$ at 90\% confidence.  Condition (i), however,
puts no constraint on halo formation time, as the $\zf$ distribution
is almost identical to that for all the 500,000 haloes.  Over the
whole mass range explored here, the fraction of haloes that meet
Condition (i) is $12.1$\%.  Condition (ii) favors a much higher Milky
Way mass, with $M_{12} > 0.6$ at 90\% confidence.  The existence of
two Magellanic Clouds results in a slight preference of younger
haloes, with $0.25 < \zf < 1.4$ at 90\% confidence, compared to $0.4 <
\zf < 1.82$ for the full sample of 500,000 haloes.  Overall, $\sim
7.4\%$ of the haloes in the sample have two Magellanic Cloud-sized
satellites, consistent with the result of Liu \etal (2011). Finally,
Condition (iii) largely recovers the lower Milky Way mass of Condition
(i), while maintaining the preference for a late formation time.  The
preferred halo mass is at $M_{12} \simeq 0.6$, with $0.25 < M_{12} <
1.4$ and $0.1 < \zf < 1.4$ at 90\% confidence.  Only 0.66\% of all the
haloes have a gap feature as required by Condition (iii), consistent
with the results of Cautun \etal (2014) and Jiang \& van den Bosch
(2015).


\section{Summary}
\label{Sec:Summary}

We have used a combination of $N$-body simulations and a
semi-analytical model to study the statistics of dark matter
substructure.  In particular, we focused on the halo-to-halo variance
of the subhalo mass fraction ($\fsub$) and the detailed shape of the
subhalo occupation distribution, and examined in detail how these two
aspects of subhalo statistics depend on the formation redshift of the
host halo. Our results can be summarized as follows.

\begin{itemize}

\item Subhalo disruption is omnipresent in $N$-body simulations.
  Based on our semi-analytical model, which is calibrated to match
  disruption statistics in the Bolshoi simulation, only $\sim40\%$
  (10\%) of subhaloes with $\macc > 10^{-4}M_0$ accreted at $\zacc=1$
  (2) survive to the present. More massive subhaloes are more likely
  to disrupt, and as many as 80\% (95\%) of all subhaloes with $\macc
  > 0.01M_0$ ($\macc > 0.1M_0$) have been disrupted since $\zacc = 1$.
  Roughly $20\%$ of subhaloes with $\macc > 10^{-4} M_0$ disrupt
  during their first orbital period. This fraction increases to $30\%$
  and $60\%$ for subhaloes with $0.01 \ge \macc < 0.1 M_0$, and $\macc
  \ge 0.1 M_0$, respectively.
  
\item The average mass fraction of first order subhaloes, $\langle
  \fsub(>10^{-4}M_0) \rangle$, scales with host halo mass in a way
  that is well approximated by $\log\langle \fsub \rangle= - 1.12 +
  0.12\log\left(M_0/10^{12}\Msunh\right)$.  The mass fraction in
  second-order subhaloes (i.e., sub-subhaloes) is roughly $\sim10\%$
  lower and well approximated by $\log\langle \fsub \rangle=- 2.26 +
  0.17 \log\left(M_0/10^{12}\Msunh\right)$.  The halo mass dependence
  of $\langle \fsub \rangle$ is the outcome of the competition between
  the accretion of new subhaloes and the evolution of existing ones,
  and therefore closely related to halo formation redshift. To good
  approximation $\log\langle \fsub \rangle= a + b\log\zf$ with $(a,b)
  = (-0.48,-2.1)$ and $(-1.2,-4.0)$ for first- and second-order
  subhaloes, respectively.  Most importantly, at fixed $\zf$, the
  average mass fraction is almost independent of $M_0$.

\item Recently, weak and strong lensing measurements have put
  constraints on the subhalo mass fraction in the Coma cluster (Okabe
  \etal 2014) and in a sample of 11 early-type galaxies from the SLACS
  survey (Vegetti \etal 2014).  We find that, in both cases, the
  lensing measurements are substantially higher than the median model
  prediction.  However, since the halo-to-halo variance is large, we
  find that the results fall within the 95 percentiles predicted for a
  $\Lambda$CDM cosmology. We caution, though, that the model is
  calibrated against dark matter-only simulations. Baryonic processes
  may affect the global subhalo mass fraction, as well as the radial
  distribution of subhaloes, especially in the central regions of host
  halos. If baryonic processes have a tendency to {\it lower} the
  amount of substructure in the central regions, as suggested by
  several recent studies (Despali \& Vegetti 2016; Fiacconi \etal
  2016; Zhu \etal 2016), the lensing measurements are likely to be
  in tension with $\Lambda$CDM predictions. While detailed
  hydrodynamical simulations are required for a fair assessment, our
  results at least highlight the importance of properly accounting for
  the halo-to-halo variance.

\item In agreement with numerous previous studies (e.g.,
  Boylan-Kolchin \etal 2010; Busha \etal 2011; Wu \etal 2013), we find
  that the subhalo occupation distribution (sHOD), $P(N|M_0)$, is
  super-Poissonian whenever $\Nave \gta 2$. However, we also find
  $P(N|M_0)$ to be clearly sub-Poissonian when $\Nave$ is small, a
  trend that had been hinted at in previous studies (e.g., Kravtsov
  \etal 2004, BK10), but not discussed in any detail. The
  sub-Poissonity is stronger if subhaloes are selected by $\macc$ or
  $\vacc$ rather than by their present mass or $\vmax$.  Based on
  these findings, we present fitting function (Eq.\ref{Eq:sigsigP})
  for the halo-to-halo scatter $\sig=\sqrt{{\rm Var}(N)}$, as a
  function $\Nave$.  We caution that there is a hint that different
  simulations do not converge regarding the deviations from
  Poissonity, with higher-resolution simulations resulting in a sHOD
  that is closer to a Poisson distribution (at $\Nave \gg 1$).

\item The 1-halo term of the galaxy two-point correlation function, or
  equivalently the galaxy power spectrum, depends on the number of
  satellite pairs $\langle N(N-1) | M_0 \rangle = \alpha^2(M_0)
  \langle N|M_0 \rangle$. Here $\alpha(M_0)$ simply expresses the
  relation between the first and second moments of the halo occupation
  distribution, $P(N|M_0)$. It is common to assume that $P(N|M_0)$ is
  a Poisson distribution, so that $\alpha(M_0) = 1.0$. However, using
  our simulation results we show that $\alpha(M_0)$ transits from
  $\sim 1.02$ at large $\Nave$ to smaller than unity at $\Nave \lta
  2$.  We demonstrate that ignoring this non-Poissonity results in
  systematic errors in the galaxy-power spectrum (and thus also in the
  corresponding two-point correlation function) of up to $\sim 3$
  percent, and with a complicated scale- and
  luminosity-dependence. Although small, such errors are a serious
  impediment for using galaxy clustering to do precision
  (i.e. percent-level) cosmology .

\item At fixed halo mass, earlier-forming haloes exhibit subhalo
  occupation statistics that are closer to Poissonian than for their
  later-forming counterparts.  This suggests that the non-Poissonity
  of subhalo occupation statistics is imprinted at accretion (i.e., in
  the mass assembly of the host halo), and that subsequent dynamical
  evolution processes (tidal stripping and heating) drive the
  occupation statistics to Poissonian.

\item We find no evidence to support the claim by Mao \etal 2015 that
  the super-Poissonian nature of subhalo occupation statistics arises
  purely from halo-to-halo variance in formation histories. In
  particular, we find that the sHOD at fixed halo mass and fixed
  formation redshift (a proxy for formation history) is not
  Poissonian, contrary to the model proposed by Mao \etal.  Although
  selecting subhaloes based on both mass and formation redshift
  slightly reduces the super-Poissonity at large $\Nave$, it increases
  the level of sub-Poissonity at small $\Nave$. We argue that the
  Poissonian results obtained by Mao \etal (2015) arise from
  randomizing a too small fraction (less than 2 percent by mass) of
  the host halo's Lagrangian volume.

\item The abundance and $\vmax$ distribution of massive satellites
  only puts loose constraints on the mass and formation redshift of
  the host halo.  In the case of the Milky Way, the requirement that
  there are three massive satellites with $\vmax \ga 30\kms$ favors a
  halo mass of $M_0 \sim 10^{11.7 \pm 0.4}\Msunh$ (90\% CL) but has
  little to no constraining power regarding its formation redshift.
  The presence of two Magellanic Clouds favors a high Milky Way halo
  mass, $M_0 \ga 10^{11.8}\Msunh$ (90\% CL), and a relatively late
  formation redshift.  Combining these constraints, and demanding a
  `gap' in the $\vmax$ distribution between that of the SMC and the
  third largest satellite galaxy, favors a low-mass, late-forming
  halo, with $0.25 < M_0/(10^{12}\Msunh) < 1.4$ and $0.1 < \zf < 1.4$,
  both at 90\% confidence.

\end{itemize}
%


\section*{Acknowledgments}

We are grateful to Yao-Yuan Mao and Andrew Hearin for valuable
discussions, to Heidi Wu for sharing her results from the Rhapsody
simulations, to all the people responsible for the Bolshoi, MultiDark,
and ELVIS simulations for making their halo catalogs publicly
available, and to the organizers and attendees of the Lorentz Center
workshop on `Dark Matter on the Smallest Scales', for creating a
productive environment that had an important impact on this
study. FvdB is supported by the Klaus Tschira Foundation and by the US
National Science Foundation through grant AST 1516962.




\label{lastpage}



\begin{thebibliography}{}

\bibitem[]{Anderhalden13}
Anderhalden D., Diemand J., 2013, \jcap, 4, 9

\bibitem[]{Auger09}
Auger M. W., Treu T., Bolton A. S., Gavazzi R., Koopmans L. V. E., Marshall P. J., Bundy K., Moustakas L. A., 2009, \apj, 705, 1099

\bibitem[]{Behroozi13a}
Behroozi P. S., Wechsler R. H., Wu H.-Y., Busha M. T., Klypin A. A., Primack J. R., 2013a, \apj, 763, 18

\bibitem[]{Behroozi13b}
Behroozi P. S., Wechsler R. H., Wu, H.-Y., 2013b, \apj, 762, 109

\bibitem[]{Benson00}
Benson A. J., Baugh C. M., Cole S., Frenk C. S., Lacey C. G., 2000, \mnras, 316, 107

\bibitem[]{BW02}
Berlind A. A., Weinberg D. H., 2002, \apj, 575, 587

\bibitem[]{Berlind03}
Berlind A. A., Weinberg D. H., Benson A. J., Baugh C. M., Cole S., Dav\'e R., Frenk C. S., Jenkins A., \etal, 2003, \apj, 593, 1

\bibitem[]{Bolton06}
Bolton A. S., Burles S., Koopmans L. V. E., Treu T., Moustakas L. A., 2006, \apj, 638, 703

\bibitem[]{Boylan-Kolchin10}
Boylan-Kolchin M., Springel V., White S.D.M., Jenkins A., 2010, \mnras, 406, 896 [BK10]

\bibitem[]{Boylan-Kolchin11}
Boylan-Kolchin M., Bullock J. S., Kaplinghat M., 2011, \mnras, 415, L40

\bibitem[]{Bose16}
Bose S., Hellwing W. A., Frenk C. S., Jenkins A., Lovell M. R., Helly J. C., Li B., Gao L., 2016, preprint (arXiv:1604.07409)

\bibitem[]{BN98}
Bryan G. L., Norman M. L., 1998, \apj, 495, 80

\bibitem[]{Bus11}
Busha M.T., Wechsler R.H., Behroozi P.S., Gerke B.F., Klypin A.A., Primack J.R., 2011, \apj, 743, 117

\bibitem[]{Cacciato13}
Cacciato M., van den Bosch F. C., More S., Mo H., Yang X., 2013, \mnras, 430, 767

\bibitem[]{Calore15}
Calore F., Cholis I., Weniger C., 2015, \jcap, 1503, 38

\bibitem[]{Cautun14}
Cautun M., Frenk C. S., van de Weygaert R., Hellwing W. A., Jones B. J. T., 2014, \mnras, 445, 2049

\bibitem[]{CS02}
Cooray A., Sheth R., 2002, PhR, 372, 1

\bibitem[]{Correa15}
Correa C. A., Wyithe J. S. B., Schaye J., Duffy A. R., 2015, \mnras, 452, 1217

\bibitem[]{DK02}
Dalal N., Kochanek C. S., 2002, \apj, 572, 25

\bibitem[]{DL04}
De Lucia G., Kauffmann G., Springel V., White S. D. M., Lanzoni B., Stoehr F., Tormen G., Yoshida N., 2004, \mnras, 348, 333

\bibitem[]{Desp16}
Despali G., Vegetti S., 2016, preprint (arXiv:1608.06938)

\bibitem[]{Dooley14}
Dooley G. A., Griffen B. F., Zukin P., Ji A. P., Vogelsberger M., Hernquist L. E., Frebel A., 2014, \apj, 786, 50

\bibitem[]{Fiacconi16}
Fiacconi D., Madau P., Potter D., Stadel J., 2016,\mnras, 824, 144

\bibitem[]{Gao05}
Gao L., Springel V., White S. D. M., 2005, \mnras, 363, L66

\bibitem[]{Gao07}
Gao L., White S. D. M., 2007, \mnras, 377, L5

\bibitem[]{Gao12}	
Gao L., Navarro J.F., Frenk C.S., Jenkins A., Springel V., White S.D.M., 2012, \mnras, 425, 2169

\bibitem[]{GK14a}
Garrison-Kimmel S., Boylan-Kolchin M., Bullock J. S., Lee K., 2014a, \mnras, 438, 2578

\bibitem[]{GK14b}
Garrison-Kimmel S., Boylan-Kolchin M., Bullock J. S., Kirby E. N., 2014, \mnras, 444, 222

\bibitem[]{Ghi98}
Ghigna S., Moore B., Governato F., Lake G., Quinn T., Stadel J., 1998, \mnras, 300, 146

\bibitem[]{Giocoli10a}
Giocoli C., Tormen G., Sheth R. K., van den Bosch F. C., 2010a, \mnras, 404, 502

\bibitem[]{Giocoli10b}
Giocoli C., Bartelmann M., Sheth R. K., Cacciato M., 2010b, \mnras, 408, 300

\bibitem[]{GK02}
Golse G., Kneib J.-P., 2002, \aap, 390, 821

\bibitem[]{Han16}
Han J., Cole S., Frenk C. S., Jing Y., 2016, \mnras, 457, 1208

\bibitem[]{Hear16}
Hearin A.P., Behroozi P.S., van den Bosch F.C., 2016, \mnras, 461, 2135

\bibitem[]{Hezaveh16}
Hezaveh Y. D., Dalal N., Marrone D. P., Mao Y.-Y., Morningstar W., Wen D., Blandford R. D., Carlstrom J. E., \etal, 2016, \apj, 823, 37

\bibitem[]{JB14}
Jiang F., van den Bosch F. C., 2014, \mnras, 440, 193	

\bibitem[]{JB15}
Jiang F., van den Bosch F. C., 2015, \mnras, 453, 3575

\bibitem[]{JB16}
Jiang F., van den Bosch F. C., 2016, \mnras, 458, 2848 [Paper I]

\bibitem[]{Klypin99}
Klypin A., Gottl\"ober S., Kravtsov A.V., Khokhlov A.M., 1999, \apj, 516, 530 

\bibitem[]{Klypin11}
Klypin A., Trujillo-Gomez S., Primack J. R., 2011, \apj, 740, 102

\bibitem[]{Koopmans2005}
Koopmans L. V. E., 2005, \mnras, 363, 1136

\bibitem[]{Kravtsov97}
Kravtsov A. V., Klypin A. A., Khokhlov A. M., 1997, \apjs, 111, 73

\bibitem[]{Kravtsov04}
Kravtsov A. V., Berlind A. A., Wechsler R. H., Klypin A. A., Gottl\"ober S., Allgood B., Primack J. R., 2004, \apj, 609, 35

\bibitem[]{Ludlow13}
Ludlow A. D., Navarro J. F., Boylan-Kolchin M., Bett P. E., Angulo R. E., Li Ming., White S. D. M., Frenk C., Springel V., 2013, \mnras, 432, 1103L

\bibitem[]{Liu11}
Liu L., Gerke B. F., Wechsler R. H., Behroozi P. S., Busha M. T., 2011, \apj, 733, 62L

\bibitem[]{Maccio08}
Macci\`o A. V., Dutton A. A., van den Bosch F. C., 2008, \mnras, 391, 1940

\bibitem[]{Mao15}
Mao Y-Y., Williamson M., Wechsler R. H., 2015, \apj, 810, 21

\bibitem[]{McBride09}
McBride J. Fakhouri O., Ma C.-P., 2009, \mnras, 398, 1858

\bibitem[]{Moore}
Moore B,. Katz N., Lake G., 1996, \apj, 457, 455  

\bibitem[]{Moster13}
Moster B. P., Naab T., White S. D. M., 2013, \mnras, 428, 3121

\bibitem[]{NFW}
Navarro J.F., Frenk C.S., White S.D.M., 1997, \apj, 490, 493

\bibitem[]{Nei08}
Neistein E., Dekel A., 2008, \mnras, 383, 615
  
\bibitem[]{Neto07}
Neto A. F., Gao L., Bett P., Cole S., Navarro J. F., Frenk C. S., White S. D. M., Springel V.,  \etal, 2007, \mnras, 381, 1450

\bibitem[]{Nierenberg14}
Nierenberg A. M., Treu T., Wright S. A., Fassnacht C. D., Auger M. W., 2014, \mnras,42, 2434

\bibitem[]{Okabe10}
Okabe N., Okura Y., Futamase T., 2010, \apj, 713, 291

\bibitem[]{Okabe14}
Okabe N., Futamase T., Kajisawa M., Kuroshima R., 2014, \apj, 784, 90

\bibitem[]{Parkinson08}
Parkinson H., Cole S., Helly J., 2008, \mnras, 383, 557

\bibitem[]{Penarrubia10}
Pe\~narrubia J., Benson A. J., Walker M. G., Gilmore G., McConnachie A. W., Mayer L., 2010, \mnras, 406, 1290

\bibitem[]{Porciani04}
Porciani C., Magliocchetti M., Norberg P., 2004, \mnras, 355, 1010

\bibitem[]{Prada12}
Prada F., Klypin A.A., Cuesta A.J., Betancort-Rijo J.E., Primack J., 2012, \mnras, 423, 3018

\bibitem[]{Purcell12}
Purcell C.W., Zentner A.R., 2012, JCAP, 12, 7

\bibitem[]{Scoccimarro01}
Scoccimarro R., Sheth R. K., Hui L., Jain B., 2001, \apj, 546, 20

\bibitem[]{Seljak00}
Seljak U., 2000, \mnras, 318, 203 

\bibitem[]{Shaw06}
Shaw L. D., Weller J., Ostriker J. P., Bode P., 2006, \apj, 646, 815

\bibitem[]{Springel01}
Springel V., White M., Hernquist L., 2001, \mnras, 549, 681

\bibitem[]{Springel08}
Springel V., Wang J., Vogelsberger M., Ludlow A., Jenkins A., Helmi A., Navarro J. F., Frenk C. S., White S. D. M., 2008, \mnras, 391, 1685

\bibitem[]{TB04}
Taylor J. E., Babul A., 2004, \mnras, 348, 811

\bibitem[]{Tinker08}
Tinker J. L., Kravtsov A. V., Klypin A., Abazajian K., Warren M., Yepes G., Gottl\"ober S, Holz D. E., 2008, \apj, 688, 709

\bibitem[]{Tinker10}
Tinker J. L., Robertson B. E., Kravtsov A. V., Klypin A., Warren M. S., Yepes G., Gottl\"ber S.,  2010, \apj, 724, 878

\bibitem[]{Tor97}
Tormen G., Bouchet F.R., White S.D.M., 1997, \mnras, 286, 865

\bibitem[]{vdB05}
van den Bosch F.C., Tormen G., Giocoli C., 2005, \mnras, 359, 1029

\bibitem[]{vdB13}
van den Bosch F. C., More S., Cacciato M., Mo H., Yang X., 2013, \mnras, 430, 725

\bibitem[]{vdB14}
van den Bosch F. C., Jiang F., Hearin A., Campbell D., Watson D., Padmanabhan N., 2014, \mnras, 445, 1713

\bibitem[]{BJ16}
van den Bosch F. C., Jiang F., 2016, \mnras, 458, 2870  [Paper II]

\bibitem[]{vdB16}
van den Bosch F. C., Jiang F., Campbell D., Behroozi P., 2016, \mnras, 455, 158

\bibitem[]{VK09}
Vegetti S., Koopmans L. V. E., 2009, \mnras, 400, 1583

\bibitem[]{Vegetti10}
Vegetti S., Koopmans L. V. E., Bolton A., Treu T., Gavazzi, R., 2010, \mnras, 408, 1969

\bibitem[]{Veg12}
Vegetti S., Lagattuta D.J., McKean J.P., Auger M.W., Fassnacht C.D., Koopmans L.V.E., 2012, Nature, 481, 341

\bibitem[]{Vegetti14}
Vegetti S., Koopmans L. V. E., Auger M. W., Treu T., Bolton A. S., 2014, \mnras, 442, 2017

\bibitem[]{Wec02}
Wechsler R.H., Bullock J.S., Primack J.R., Kravtsov A.V., Dekel A., 2002, ApJ, 568, 52

\bibitem[]{Wechsler06}
Wechsler R. H., Zentner A. R., Bullock James. S., Kravtsov A. V., Allgood B., 2006, \apj, 652, 71

\bibitem[]{Wu13} 
Wu H.-Y., Hahn O., Wechsler R. H., Behroozi P. S., Mao Y.-Y., 2013, \apj, 767, 23

\bibitem[]{Xu15}
Xu D., Sluse D., Gao L., Wang J., Frenk C., Mao S., Schneider P., Springel V., 2015, \mnras, 447, 3189

\bibitem[]{YMB03}
Yang X., Mo H. J., van den Bosch F. C., 2003, \mnras, 339, 1057

\bibitem[]{Zhao09}
Zhao D. H., Jing Y. P., Mo H. J., B\:orner G., 2009, \apj, 707, 354

\bibitem[]{Zentner05}
Zentner A.R., Berlind A.A., Bullock J.S., Kravtsov A.V., Wechsler R.H., 2005, \apj, 624, 505

\bibitem[]{Zentner14}
Zentner A. R., Hearin A. P., van den Bosch F. C., 2014, \mnras, 443, 3044

\bibitem[]{Zheng05}
Zheng Z., Berlind A. A., Weinberg D. H., Benson A. J., Baugh C. M., Cole S.; Dav\'e R., Frenk C. S. \etal, 2005, \apj, 633, 791 

\bibitem[]{Zhu16}
Zhu Q., Marinacci F., Maji M., Li Y., Springel V., Hernquist L., 2016, \mnras, 458, 1559

\end{thebibliography}
\end{document}